\documentclass[a4paper,english,sort&compress]{revtex4-1}
\usepackage[T1]{fontenc}
\pdfoutput=1
\usepackage[latin9]{inputenc}
\usepackage{xcolor}
\usepackage{pdfcolmk}
\usepackage{babel}
\usepackage{prettyref}
\usepackage{amsmath}
\usepackage{amssymb}
\usepackage{graphicx}
\PassOptionsToPackage{normalem}{ulem}
\usepackage{ulem}
\usepackage[unicode=true,pdfusetitle,
 bookmarks=true,bookmarksnumbered=false,bookmarksopen=false,
 breaklinks=false,pdfborder={0 0 1},backref=false,colorlinks=false]
 {hyperref}

\makeatletter

\pdfpageheight\paperheight
\pdfpagewidth\paperwidth

\providecommand{\tabularnewline}{\\}
\newcommand{\lyxdot}{.}

\providecolor{lyxadded}{rgb}{0,0,1}
\providecolor{lyxdeleted}{rgb}{1,0,0}

\DeclareRobustCommand{\lyxsout}[1]{\ifx\\#1\else\sout{#1}\fi}

\makeatother

\begin{document}
	
\title{The cavity approach for Steiner trees packing Problems}
\author{Alfredo Braunstein}
\email{alfredo.braunstein@polito.it}

\address{DISAT, Politecnico di Torino, Corso Duca Degli Abruzzi 24, Torino,
Italy}
\address{Italian Institute for Genetic Medicine (form. HuGeF), Via Nizza 52, Torino, Italy}
\address{INFN Sezione di Torino, Via P. Giuria 1, I-10125 Torino, Italy}
\address{Collegio Carlo Alberto, Via Real Collegio 1, Moncalieri, Italy}
\author{Anna Paola Muntoni}
\email{anna.muntoni@polito.it}
\address{Laboratoire de physique th$\acute{e}$orique, D$\acute{e}$partement de physique de l'ENS, $\acute{E}$cole normale sup$\acute{e}$rieure, PSL University, Sorbonne Universit$\acute{e}$, CNRS, 75005 Paris, France}
\address{DISAT, Politecnico di Torino, Corso Duca Degli Abruzzi 24, Torino,
Italy}
\begin{abstract}
The Belief Propagation approximation, or cavity method, has been recently
applied to several combinatorial optimization problems in its zero-temperature
implementation, the max-sum algorithm. In particular, recent developments
to solve the edge-disjoint paths problem and the prize-collecting
Steiner tree problem on graphs have shown remarkable results for several
classes of graphs and for benchmark instances. Here we propose a generalization
of these techniques for two variants of the Steiner trees \textit{packing}
problem where multiple ``interacting'' trees have to be sought within
a given graph. Depending on the interaction among trees we distinguish
the Vertex-Disjoint Steiner trees Problem, where trees cannot share
nodes, from the Edge-Disjoint Steiner trees Problem, where edges cannot
be shared by trees but nodes can be members of multiple trees. Several
practical problems of huge interest in network design can be mapped
into these two variants, for instance, the physical design of Very
Large Scale Integration (VLSI) chips.

The formalism described here relies on two components edge-variables
that allows us to formulate a massage-passing algorithm for the V-DStP
and two algorithms for the E-DStP differing in the scaling of the
computational time with respect to some relevant parameters. We will
show that through one of the two formalisms used for the edge-disjoint
variant it is possible to map the Max-Sum update equations into a
weighted maximum matching problem over proper bipartite graphs. We
developed a heuristic procedure based on the Max-Sum equations that
shows excellent performance in synthetic networks (in particular outperforming
standard multi-step greedy procedures by large margins) and on large
benchmark instances of VLSI for which the optimal solution is known,
on which the algorithm found the optimum in two cases and the gap
to optimality was never larger than 4\%.

\tableofcontents{}
\end{abstract}
\maketitle

\section{Introduction}

The minimum Steiner tree problem (MStP) is an important combinatorial
problem that consists in finding a connected sub-graph within a given
weighted graph, able to span a subset of vertices (\textit{\emph{called}}\textit{
terminal}s) with minimum cost. It is easy to see that if weights are
strictly positive the sub-graph satisfying all these constraints must
be a tree.

The decisional problem of determining whether a solution within a
given cost bound exists is NP-complete (it was one of Karp's original
21 problems \citep{Karp1972}). The large difficulty
of the MStP can be seen to arise from the large space of subsets of
non-terminal vertices (\textit{Steiner nodes}). There exist several
variants and generalizations of the MStP; one of the most studied
is the prize-collecting Steiner problem (PCStP) that have many applications
in network technologies, such as optimal heating and optical fibers
distribution \citep{ljubic_algorithmic_2005}, in biology, e.g. in
finding signal pathways in a cell \citep{bailly-bechet_finding_2011}.
In the prize-collecting variant the notion of terminals is relaxed
so that every vertex has an associated prize (or reward). The prize
of included nodes is counted negatively in the solution cost (so that
\textit{profitable vertices }\textit{\emph{with positive reward}}
lower the total cost). In this variant the cost of the optimal tree
will be the best trade-off between prizes of included nodes and the
cost of their connections given by edge-weights. 

In this work we will address the \emph{packing of Steiner Trees} problem
where we aim at finding, within the same graph, multiple Steiner trees
which span disjoints sets of terminals in its original and prize-collecting
versions. We consider two different variants regarding the interaction
among trees. In the Vertex-disjoint Steiner trees problem (V-DStP),
different trees cannot share vertices (and consequently they cannot
share edges either); in the Edge-disjoint Steiner tree problem (E-DStP)
only edge sets are pairwise disjoint but nodes can be shared by different
trees. Being generalizations of PCStP, both problems are NP-hard;
from a practical point of view the packing problems are more difficult
than their single-tree counterpart as it can be seen from the fact
that even finding feasible solutions, i.e. trees satisfying the interaction
constraints (regardless the cost), is NP-hard \citep{hoang_steiner_2012}.
In addition to its mathematical interest, a lot of attention is devoted
to the practical solution of packing of Steiner trees problems since
several layout design issues arising from Very Large Scale Integration
(VLSI) circuits \citep{cohoon_beaver:_1988,burstein_hierarchical_1983,luk1985greedy}
can be mapped into these variants of the MStP \citep{Grotschel1997, hoang_steiner_2012}.
Integrated systems are composed by a huge number of logical units,
called cells, typically arranged in 2D or 3D grids. Some specific
cells, forming the so-called \textsl{net}, must be connected to one
another in order to satisfy some working conditions. The physical
design phase of these circuits addresses the problem of connecting
each element of a net minimizing some objective function, namely the
power consumption induced by the wires of the connection. It can be
easily seen that connecting the cells of a net at minimum power consumption
is equivalent to solve a MStP on a 2D or 3D grid graph. Thus, the
problem of concurrently connecting multiple and disjoint sets of nets
can be easily mapped into a V-DStP or an E-DStP. The most common approaches
to these combinatorial optimization problems rely on linear programming
formulations, for instance, the multi-commodity flow model \citep{hoang_steiner_2012}.

In this work we devise three different models to represent these two
problems: one for V-DStP and two for E-DStP, the first one more suitable
for graphs where the density of terminals is low and the second for
instances with low graph connectivity. We attempt their solution through
the Cavity Method of statistical physics and its algorithmic counterpart,
the Belief Propagation (BP) iterative algorithm (or rather, its zero-temperature
limit, the Max-Sum (MS) algorithm \citep{mezard_information_2009,mezard_cavity_2003}).
This technique is an approximation scheme first developed to study
disordered systems and nowadays applied to a wide range of optimization
problems. Once a proper set of variables is defined, the optimization
problem, i.e. the constrained minimization of a cost function, can
be mapped into the problem of finding the ground-state(s) of a generalized
system with local interactions. Ground-states can be investigated
through observables related to the Boltzmann-Gibbs distribution at
zero temperature but, in most of the interesting cases, their exact
evaluation involves impractical computations. MS consists in iterating
closed massage-passing equations on a factor graph, closely related
to the original graph, that, at convergence, provide an estimate of
the marginal probability distribution of the variables of interest. 

The cavity method can be proven to be exact on tree graphs (and also
in some models on random networks in the asymptotic limit) but nevertheless
in practice reaches notable performances on arbitrary graphs. It should
be noted that in a simplified version of the problem (the minimum
spanning tree), fixed points of Max-Sum can be proven to parametrize
the optimal solution \citep{bayati_rigorous_2008}. As usual, the
iterative solution of the Max-Sum equations involve the solution of
a related problem in a local star-shaped sub-graph which for some
of these models is not trivial (i.e. its naive solution is exponentially
slow in the degree). We devise a mapping of the problem into a minimum
matching problem that can be solved in polynomial time in the degree
(leading e.g. to linear time per iteration on Erd\H{o}s--Rényi random
graphs).

In combination with these three initial models, a variant called the
\emph{flat }formalism, borrowed from \citep{braunstein_practical_2016},
can be independently included, leading to six different model combinations
for the two problems. The flat formalism is more suitable for graphs
with large diameter and/or few terminals as it allows to reduce considerably
the solution space. Interestingly, the resulting \emph{flat} models
can be seen as generalizations of both \citep{braunstein_practical_2016}
and \citep{altarelli_edge-disjoint_2015,bacco_shortest_2014} as the
edge-disjoint path problems in the last two publications can be seen
as a packing of Steiner trees problem in which each tree has exactly
two terminals. 

With these algorithmic tools on hand, we perform numerical simulations
of complete, Erd\H{o}s--Rényi and random regular graphs and on benchmark
instances of V-DStP arising from the VLSI design problem.

\section{Two Steiner packing problems \label{sec:Definition}}

Given a graph $G=\left(V,\,E\right)$ whose vertices have non-negative
real prizes $\left\{ c_{i}^{\mu}:i\in V,\mu=1,\ldots,M\right\} $
and whose edges have real positive weights $\left\{ w_{ij}^{\mu}:\left(i,j\right)\in E,\mu=1,\ldots,M\right\} $,
we consider the problem of finding $M$ connected sub-graphs $G_{\mu}=\left(V_{\mu},E_{\mu}\right)$
spanning disjoint sets of terminals $\left\{ T_{\mu}\subseteq V_{\mu},\mu=1,\ldots,M\right\} $
that minimize the following cost or energy function
\begin{equation}
H=\sum_{\mu}\left[\sum_{i\in V\setminus V_{\mu}}c_{i}^{\mu}+\sum_{(i,j)\in E_{\mu}}w_{ij}^{\mu}\right]\label{eq:ham}
\end{equation}

This definition of the cost is extremely general: node prizes and
edge costs can depend on sub-graph $\mu$. For directed graphs, we
can admit $w_{ij}^{\mu}\neq w_{ji}^{\mu}$ by considering oriented
trees (the trees we will consider will be ultimately rooted and thus
oriented). In the following we refer to vertices with strictly positive
prizes as (generalized) terminals in analogy with the MStP. This particular
case can be integrated in our formalism imposing $c_{i}^{\mu}=+\infty$
if node $i$ is a (true) terminal of tree $\mu$ (it suffices to have
a large enough value for $c_{i}^{\mu}$ instead of $+\infty$), and
$c_{j}^{\mu}=0,\,\forall\mu$ for any non-terminal node $j\in V$.
Since we interpret the solution-trees as networks that allow terminals
to ``communicate'' we will refer to each sub-graph $G_{\mu}$ as
a ``communication'' $\mu$ flowing within the graph.

Subsets $G_{\mu}$ must satisfy some interaction constraints depending
on the packing variant we are considering. In the \emph{Vertex-disjoint
Steiner trees Problem} (V-DStP), vertex-sets $V_{\mu}$ must be pairwise
disjoint, i.e. $V_{\mu}\cap V_{\nu}=\emptyset$ if $\mu\neq\nu$ and,
consequently, also edge sets will be pairwise disjoint. In the \emph{Edge-disjoint
Steiner trees Problem} (E-DStP), only edge sets must be pairwise disjoints,
i.e. $E_{\mu}\cap E_{\nu}=\emptyset$ if $\mu\neq\nu$, but vertex
sets can overlap. 

\section{An arborescent representation \label{sec:Representation}}

To deal with these two combinatorial optimization problems we will
define a proper set of interacting variables defined on a factor graph
which is closely related to the original graph $G$. The factor graph
is the bipartite graph of factors (or compatibility functions) and
variables, in which an edge between a factor and a variable exists
if the function depends on the variable. More precisely, to each vertex
$i\in V$ we associate a factor node $\psi_{i}$ and to each edge
$\left(i,j\right)\in E$ we associate a two components variable $\left(d_{ij},\mu_{ij}\right)\in\left\{ -D,\dots,0,\dots,D\right\} \times\left\{ 0,\dots,M\right\} $.
Our choice of the edge-variables is similar to the one adopted in
\citep{braunstein_practical_2016} but here, in addition to a ``depth''
component, we introduce a ``communication'' variable $\mu_{ij}$
by which we label edges forming different trees. 

Compatibility functions $\psi_{i}$ are defined in a way that allowed
configurations of variables $\left(\boldsymbol{d},\boldsymbol{\mu}\right)\doteq\left\{ \left(d_{ij},\mu_{ij}\right):\left(i,j\right)\in E\right\} $
are in one to one correspondence to feasible solutions of the Vertex-disjoint
or Edge-disjoint variant of the Steiner trees problem. In particular,
in order to ensure Steiner sub-graphs to be trees, i.e. to be connected
and acyclic, we impose local constraints on variables $\boldsymbol{d}_{i}=\left\{ d_{ij}:j\in\partial i\right\} $
and $\boldsymbol{\mu}_{i}=\left\{ \mu_{ij}:j\in\partial i\right\} $
through compatibility functions $\psi_{i}\left(\boldsymbol{d}_{i},\boldsymbol{\mu}_{i}\right)$
that will be equal to one if the constraints are satisfied or zero
otherwise. 

Consider a solution to the V-DStP or the E-DStP. Each variable $\mu_{ij}$
takes value from the set $\left\{ 0,1,\ldots,M\right\} $ and denotes
to which sub graph, if any, does the edge $\left(i,j\right)$ belongs;
the state $\mu_{ij}=0$ will conventionally mean that no tree employs
the edge $\left(i,j\right)$. Components $d_{ij}\in\left\{ -D,\ldots,0,\ldots,D\right\} $
have a meaning of ``depth'' or ``distance'' within the sub-graph.
Value $d_{ij}=0$ conventionally means that such edge is not employed
by any communication and thus it is admitted if and only if the associated
$\mu_{ij}=0$.

Being the interactions among nodes different as we deal with the V-DStP
or the E-DStP, we will define two different compatibility functions,
$\psi_{i}^{V}$ and $\psi_{i}^{E}$, for the two problems. Both functions
will be written with the help of a single-tree compatibility function
$\psi_{i}^{\mu}$ for two different formulations of the constraints,
the \textit{branching} and the \textit{flat} model.

\subsubsection{Branching model}

Let us consider a sub-graph $G_{\mu}$ constituting part of the solution
for the V-DStP or the E-DStP. For each node $i\in V_{\mu}$, the variable
$d_{ij}$ measures the length, in ``steps'', of the unique path
from node $i$ to root $r_{\mu}$ passing through $j\in\partial i$.
Variable $d_{ij}$ will be strictly positive (negative) if $j$ is
one step closer (farther) than $i$ to root $r_{\mu}$. Thus, every
edge will satisfy the anti-symmetric condition $d_{ij}=-d_{ji}$ and
$\mu_{ij}=\mu_{ji}$. A directed tree structure is guaranteed if,
mathematically, the following single-tree compatibility function

\begin{equation}
\psi_{i}^{\mu,\,b}\left(\boldsymbol{d}_{i},\boldsymbol{\mu}_{i}\right)=\prod_{j\in\partial i}\delta_{\mu_{ji},0}\delta_{d_{ji},0}+\sum_{d>0}\sum_{j\in\partial i}\left[\delta_{\mu,\mu_{ji}}\delta_{d_{ji},-d}\prod_{k\in\partial i\setminus j}\left(\delta_{\mu,\mu_{ki}}\delta_{d_{ki},d+1}+\delta_{\mu_{ki},0}\delta_{d_{ki},0}\right)\right]\label{eq:compatib-depth}
\end{equation}
equals to $1$ for every nodes in the graph. Here $\delta_{x,y}$
is the discrete Kronecker delta function equal to $1$ if $x=y$ and
$0$ otherwise. The first part of the equation describes a feasible
assignment for a node $i$ that does not participate to any communication:
its local variables satisfy $\boldsymbol{d}_{i}=\boldsymbol{0}$ and
$\boldsymbol{\mu}_{i}=\boldsymbol{0}$. The second addend considers
a second case in which $i$ is member of tree $\mu$ at distance $d$
from the root; in this case there will exist only one neighbor $j$,
one step closer to the root than $i$, such that $d_{ij}=d$ (as a
consequence $d_{ji}=-d_{ij}<0$) and $\mu_{ij}=\mu$ . All the remaining
neighbors $k\in\partial i\backslash j$ may not be members of the
communication ($\mu_{ki}=0$, $d_{ki}=0$) or being part of $G_{\mu}$
($\mu_{ki}=\mu$) as children of $i$. In the latter any $k$ is one
step further than $i$ from the root $r_{\mu}$ and thus their depths
must be increased of one unit, namely $d_{ki}=d+1$. Notice that for
a given feasible assignment of the local variables the summations
over the possible positive depths and neighbors in \eqref{eq:compatib-depth}
reduce to a single term that corresponds to the unique $j\in\partial i$
such that $d_{ji}<0$; in this case $\psi_{i}^{\mu,\,b}\left(\boldsymbol{d}_{i},\boldsymbol{\mu}_{i}\right)=1$.
Instead, for all unfeasible assignments $\psi_{i}^{\mu,\,b}\left(\boldsymbol{d}_{i},\boldsymbol{\mu}_{i}\right)=0$.

In the following this representation of the tree structure will be
referred as the \textit{branching }formalism and it will be denoted
using an apex ``b'' in its compatibility function as in \eqref{eq:compatib-depth}.

\subsubsection{The flat formalism \label{subsec:The-flat-formalism}}

The diameter of solutions representable using the \textit{branching
model} strongly depends on the value of the parameter $D$ which is
the maximum allowed distance from any leaf and the root of the tree.
A small value of the depth parameter can certainly prevent the representation
of more elongated and, possibly, more energetically favored solutions
but, at the same time, a big value of $D$ will significantly slow
down the computation of the compatibility function in \eqref{eq:compatib-depth}.
The \textit{flat} model relaxes the depth-increasing constraint in
the sense that, under certain conditions, it allows chains of nodes,
within the solution, with equal depth. According to a flat representation,
the depth variable increases of one unity if a node $i$ is a terminal
node or there exist two or more neighbors connected to $i$ within
a sub-graph $G_{\mu}$, i.e. the degree of node $i$ within the sub-graph
is more than two. It can be shown \citep{braunstein_practical_2016}
that for $D=T$, where $T=|T_{\mu}|$ is the number of terminals per
communication, these constraints admit all feasible trees plus some
extra structures which contain disconnected cycles with no terminals
and thus are energetically disfavored. This additional ``flat''
assignment applies to only non-terminal and non-branching nodes in
the solution and the corresponding compatibility function can be stated
as follows:

\begin{equation}
\psi_{i}^{\mu,\,f}\left(\boldsymbol{d}_{i},\boldsymbol{\mu}_{i}\right)=\delta_{c_{i}^{\mu},0}\sum_{\substack{d>0}
}\sum_{k\in\partial i}\delta_{\mu,\mu_{ki}}\delta_{-d,d_{ki}}\sum_{l\in\partial i\backslash k}\delta_{\mu,\mu_{li}}\delta_{d_{li},d}\prod_{m\in\partial i\backslash\left\{ k,l\right\} }\delta_{\mu_{mi},0}\delta_{d_{mi},0}\label{eq:compa-func-flat}
\end{equation}
More precisely, if node $i$ is not a terminal of sub-graph $\mu$
but it is a Steiner node at distance $D\leq d<0$, there exists one
of its neighbors $k$ (its parent within the solution) such that $d_{ik}=d$.
Node $i$ cannot be a leaf \footnote{Being a non-terminal node, the path connecting $i$ to the closest
terminal does not carry any advantage in terms of connection and only
increases the cost of the solution} and therefore there must be a child of $i$, a node $l\in\partial i\backslash k$,
at the same depth $d$. All the remaining neighbors $m\in\partial i\backslash\left\{ k,l\right\} $
must have $d_{m,i}=0$ to preserve the chain structure. As for the
compatibility function in \eqref{eq:compatib-depth}, \eqref{eq:compa-func-flat}
returns $1$ for an assignment that satisfies the ``flat'' constraints:
among all the addends only the term in which two neighbors have the
same non-zero depth (with opposite sign) will survive. For all assignments
not satisfying this property we will have $\psi_{i}^{\mu,\,f}\left(\boldsymbol{d}_{i},\boldsymbol{\mu}_{i}\right)=0$.
Finally, the single-tree compatibility function $\psi_{i}^{\mu}=1-\left(1-\psi_{i}^{\mu,\,b}\right)\left(1-\psi_{i}^{\mu,\,f}\right)$
of configuration satisfying exactly one of the two constraints can
be written as

\begin{align}
\psi_{i}^{\mu}\left(\boldsymbol{d}_{i},\boldsymbol{\mu}_{i}\right) & =\psi_{i}^{\mu,\,b}\left(\boldsymbol{d}_{i},\boldsymbol{\mu}_{i}\right)+\psi_{i}^{\mu,\,f}\left(\boldsymbol{d}_{i},\boldsymbol{\mu}_{i}\right)\label{eq:Single-Tree-comp}
\end{align}
since $\psi_{i}^{\mu,\,b}\psi_{i}^{\mu,\,f}\equiv0$.

\subsection{Constraints for the Vertex-Disjoint Steiner trees problem}

In the V-DStP a node $i$ can belong to none or at most one sub-graph
$G_{\mu}$and, as a consequence, its neighbor edges can either participate
to the same communication or be unused. For this reason we can consider
communication wise the topological constraints applied to the neighborhood
of each node.. The compatibility function $\psi_{i}^{V}$ can be then
expressed as the sum over all possible trees of a single-tree compatibility
function $\psi_{i}^{\mu}$ in \eqref{eq:Single-Tree-comp}.
\begin{eqnarray}
\psi_{i}^{V}\left(\boldsymbol{d}_{i},\boldsymbol{\mu}_{i}\right) & = & \sum_{\mu=1}^{M}\psi_{i}^{\mu}\left(\boldsymbol{d}_{i},\boldsymbol{\mu}_{i}\right)\label{eq:psi:vertexdisjoint}
\end{eqnarray}

\subsection{Constraints for the Edge-Disjoint Steiner trees problem}

Differently to the V-DStP, in the E-DStP a vertex can belong to an
arbitrary number of communications (including zero) with the constraint
that the local tree structure must be concurrently satisfied for each
communication. If the node does not participate in the solution we
must admit configurations in which $\boldsymbol{d}_{i}=\boldsymbol{0}$
if $\boldsymbol{\mu}_{i}=\boldsymbol{0}$. For the remaining cases,
if some neighbors $k\in\partial i$ is a members of a Steiner tree
$\mu$, its distances $d_{ki}$ will be different from zero if $\mu_{ki}=\mu$,
and, additionally, they will satisfy the topological constraints.
We can mathematically express such conditions through the compatibility
functions

\begin{equation}
\psi_{i}^{E,b}\left(\boldsymbol{d}_{i},\boldsymbol{\mu}_{i}\right)=\prod_{\mu=1}^{M}\left[\prod_{k\in\partial i}\delta_{d_{ki}\delta_{\mu_{ki},\mu},0}+\sum_{d>0}\sum_{k\in\partial i}\delta_{d_{ki}\delta_{\mu_{ki},\mu},-d}\prod_{l\in\partial i\setminus k}\left(\delta_{d_{li}\delta_{\mu_{li},\mu},d+1}+\delta_{d_{li}\delta_{\mu_{li},\mu},0}\right)\right]\label{eq:edge-disj-depth-inc}
\end{equation}
for the branching model and
\begin{equation}
\psi_{i}^{E,f}\left(\boldsymbol{d}_{i},\boldsymbol{\mu}_{i}\right)=\prod_{\mu}\left[\delta_{c_{i}^{\mu},0}\sum_{\substack{d>0}
}\sum_{k\in\partial i}\delta_{-d,d_{ki}\delta_{\mu_{ki},\mu}}\sum_{l\in\partial i\backslash k}\delta_{\mu,\mu_{li}}\delta_{d_{li}\delta_{\mu_{li},\mu},d}\prod_{m\in\partial i\backslash\left\{ k,l\right\} }\delta_{d_{mi}\delta_{\mu_{mi},\mu},0}\right]
\end{equation}
for the flat model. Notice that we can express 
\begin{equation}
\psi_{i}^{E}\left(\boldsymbol{d}_{i},\boldsymbol{\mu}_{i}\right)=\psi_{i}^{E,b}\left(\boldsymbol{d}_{i},\boldsymbol{\mu}_{i}\right)+\psi_{i}^{E,f}\left(\boldsymbol{d}_{i},\boldsymbol{\mu}_{i}\right)
\end{equation}
or eventually, if we define $\tilde{d_{ki}}=d_{ki}\delta_{\mu_{ki},\mu}$
, we can rephrase it as a product over single-tree compatibility functions
(compare to \eqref{eq:psi:vertexdisjoint})
\begin{eqnarray}
\psi_{i}^{E}\left(\boldsymbol{d}_{i},\boldsymbol{\mu}_{i}\right) & = & \prod_{\mu=1}^{M}\psi_{i}^{\mu}\left(\tilde{\boldsymbol{d}}_{i},\boldsymbol{\mu}_{i}\right)\label{eq:psi-edge-disj}
\end{eqnarray}

\begin{figure}[h]
\begin{centering}
\includegraphics[width=0.9\textwidth]{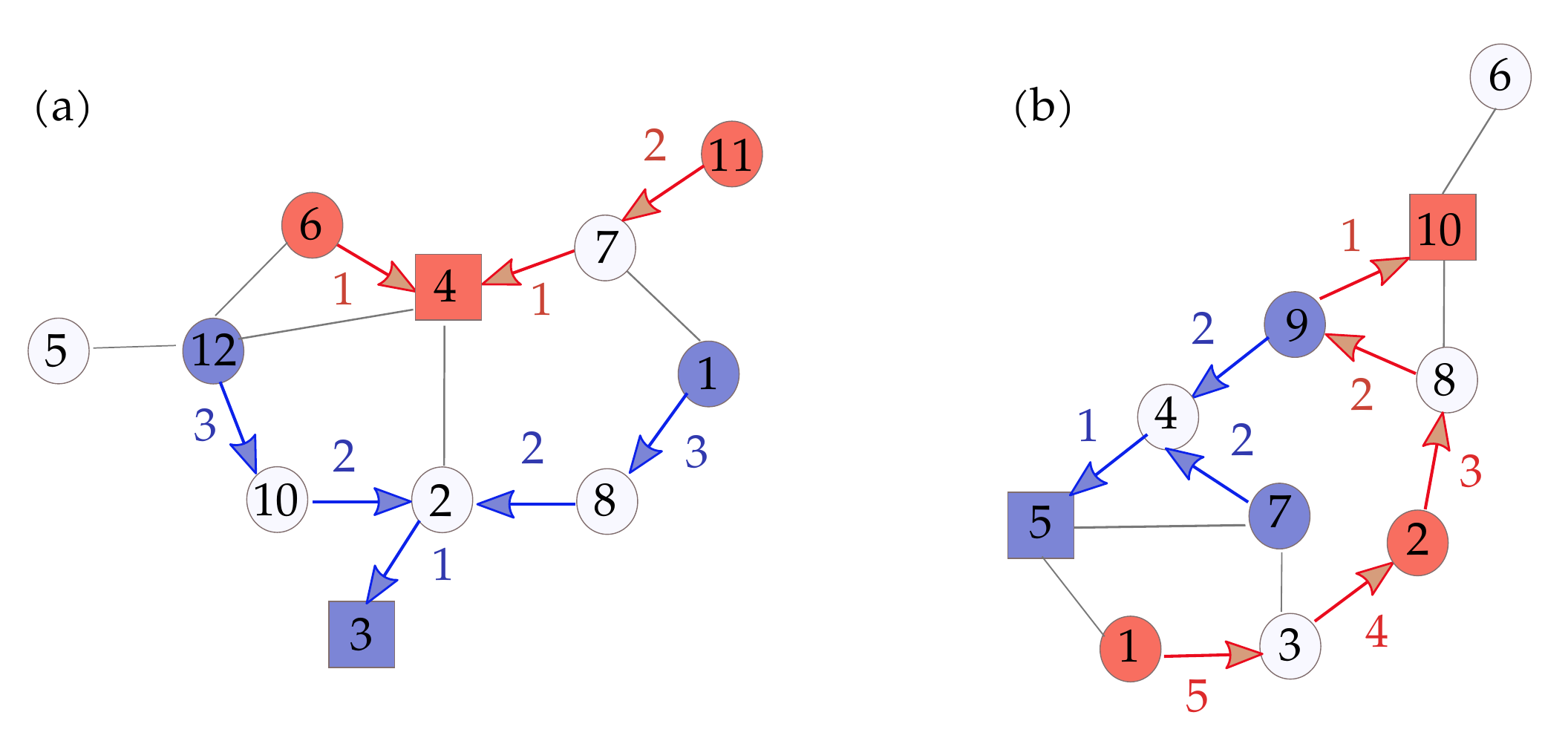}
\par\end{centering}
\caption{Figures (a) and (b) show a feasible assignment of the variables for
the V-DStP (left) and the E-DStP (right) using the branching model.
\label{fig:smallgraphs}}
\end{figure}

\begin{figure}[h]
\begin{centering}
\includegraphics[width=0.9\textwidth]{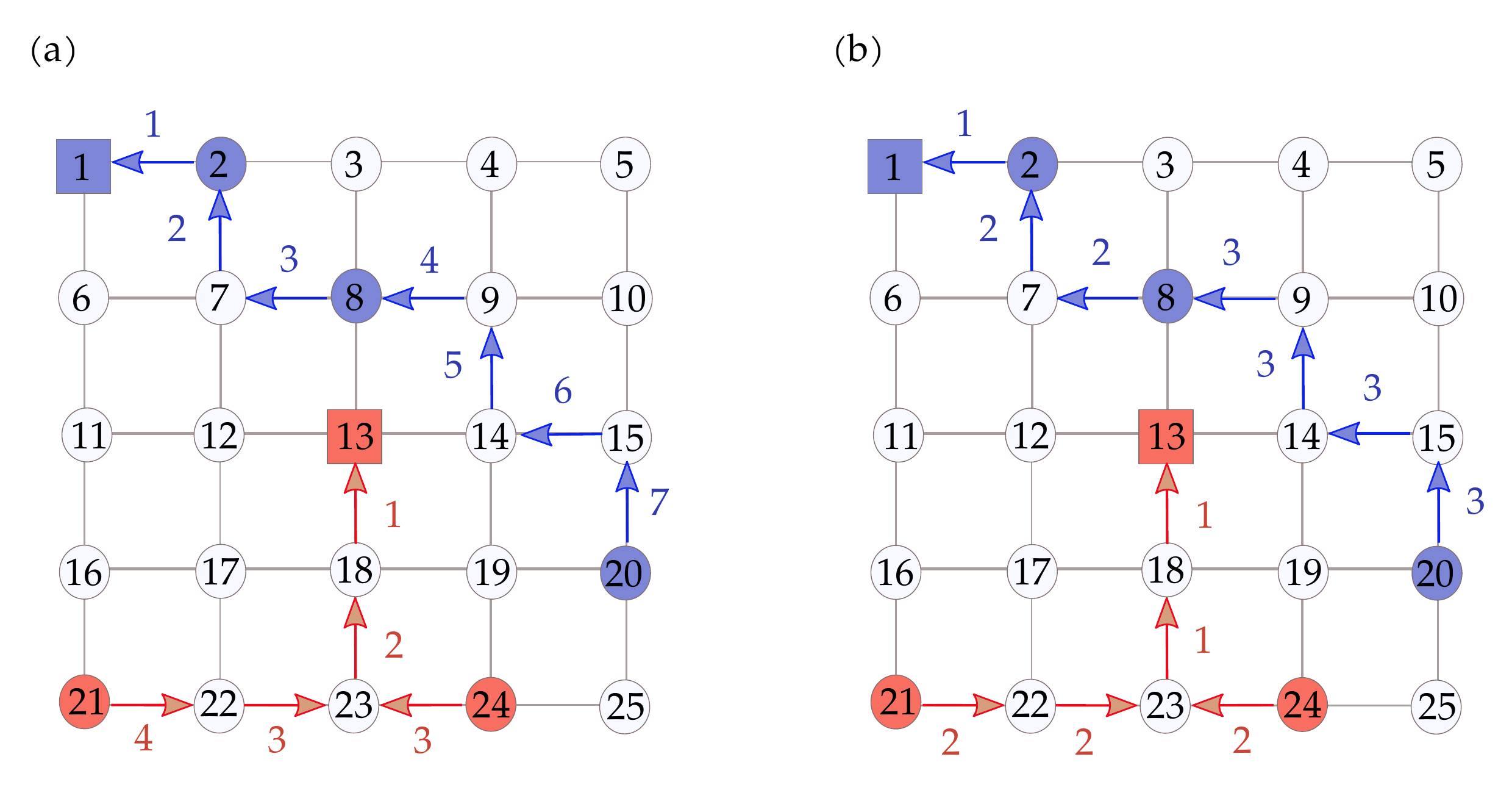}
\par\end{centering}
\caption{Figures (a) and (b) picture the same solution to a V-DStP using the
branching formalism (figure (a)) and the flat representation (figure
(b)) \label{fig:small-grids}}
\end{figure}

Some examples of feasible assignments of variables for both branching
and flat models are shown in figures \ref{fig:smallgraphs} and \ref{fig:small-grids}.
On the left (figure \ref{fig:smallgraphs} (a)) we see one instance
of the V-DStP represented through the branching formalism and containing
two sub-graphs, the ``red'' having root ``4'' and the ``blue''
rooted at node ``3''; on the right (figure \ref{fig:smallgraphs}
(b)) two ``red'' and ``blue'' edge-disjoint Steiner trees, rooted
at ``10'' and ``5'' respectively. Roots are represented as square
nodes in contrast to circle colored nodes that are terminals. Edges
employed in the solutions are figured as arrows whose labels denote
the value of the (positive) depth component while the color mirrors
the communication component. In agreement with our branching representation
we see that depth components increase as we cover the solution from
the root to the leaves for both problems. In the figure on the left,
nodes of the vertex-disjoint trees are members either of the ``red''
or the ``blue'' trees but such constraint is relaxed in the figure \ref{fig:smallgraphs}
(b) for E-DStP. In fact, we allow node ``9'' to be a terminal of
communication ``blue'' and a Steiner node of the ``red'' tree
as two incident edges, $(10,9)$ and $(9,8)$, belong to the ``red''
solution.

To underline the differences between the branching and the flat formalism,
we picture in figure \ref{fig:small-grids} (a) and (b) the same solution
to the V-DStP on a grid graph using both models. According to the
branching formalism, we see in figure \ref{fig:small-grids} (a) that
we need a minimum depth of $D=7$ to allow all terminals of the ''blue''
communication to reach the root node ``1'' . Notice that since the
tree is actually a chain of nodes, the same solution can be represented
in the flat formalism using $D=3$ as shown in figure \ref{fig:small-grids}
(b). In fact, only each time we reach a terminal node the depth variable
increases of one unit. Depth variables must increase in another condition,
precisely when we reach a branching point: this is exactly what happens
in the neighborhood of node ``23'' of the ``red'' solution in figure
\ref{fig:small-grids} (b).


\section{Boltzmann distribution and marginals \label{sec:Boltzmann-BP}}

The formalism introduced above allows us to map each solution of the
packing of Steiner trees to a certain assignment of variables $\boldsymbol{d}=\left\{ d_{ij}:\left(i,j\right)\in E\right\} $
and $\boldsymbol{\mu}=\left\{ \mu_{ij}:\left(i,j\right)\in E\right\} $
of the associated factor graph. The cost function in \eqref{eq:ham}
can be then expressed in terms of the new variables as

\begin{equation}
H\left(\boldsymbol{d},\boldsymbol{\mu}\right)=\sum_{\mu=1}^{M}\left[\sum_{i\in V}c_{i}^{\mu}\prod_{k\in\partial i}\left(1-\delta_{\mu_{ki},\mu}\right)+\sum_{\substack{d_{ij}>0:\\
\mu_{ij}=\mu
}
}w_{ij}\right]\label{eq:multi:ham}
\end{equation}
where, for sake of simplicity, we consider the ``homogeneous'' case
$w_{ij}^{\mu}=w_{ij},$ and $\psi_{i}$ can be either equal to $\psi_{i}^{V}$
or $\psi_{i}^{E}$. 

The Boltzmann-Gibbs distribution associated with the energy $H\left(\boldsymbol{d},\boldsymbol{\mu}\right)$
is given by

\begin{equation}
P_{\beta}\left(\boldsymbol{d},\boldsymbol{\mu}\right)=\frac{\prod_{i}\psi_{i}\left(\boldsymbol{d}_{i},\boldsymbol{\mu}_{i}\right)e^{-\beta H\left(\boldsymbol{d},\boldsymbol{\mu}\right)}}{Z_{\beta}}\label{eq:multi:prob}
\end{equation}
in which $\beta$ is a positive parameter, called the ``inverse temperature''
as in the statistical mechanics framework, and the normalization constant 

\[
Z_{\beta}=\sum_{\boldsymbol{d},\boldsymbol{\mu}}\prod_{i}\psi_{i}\left(\boldsymbol{d}_{i},\boldsymbol{\mu}_{i}\right)e^{-\beta H\left(\boldsymbol{d},\boldsymbol{\mu}\right)}
\]
is the partition function. Configurations of the variables that do
not satisfy the topological constraints will have zero probability
measure, whereas all other configurations will be weighted according
to the sum of weights of used edges and of the penalties of non-employed
nodes. In the limit $\beta\rightarrow+\infty$ the distribution will
be concentrated in the configuration that minimizes $H\left(\boldsymbol{d},\boldsymbol{\mu}\right)$
(the \emph{ground state }of the system) that are exactly the solutions
of the optimization problems. As a consequence also the marginal probability
densities, for $\beta\rightarrow+\infty$, associated with each edge

\begin{equation}
P_{ij}^{\beta\rightarrow+\infty}\left(d_{ij},\mu_{ij}\right)=\sum_{\bar{\boldsymbol{d}},\bar{\boldsymbol{\mu}}}P_{\beta\rightarrow+\infty}\left(\boldsymbol{d},\boldsymbol{\mu}\right)\delta_{d_{ij},\bar{d}_{ij}}\delta_{\mu_{ij},\bar{\mu}_{ij}}\quad\forall\,\left(i,j\right)\in E\label{eq:multi:marg}
\end{equation}
will be concentrated and therefore it suffices to estimate them to
gain knowledge about the optimal configuration. Thus our assignment
of the variables will be given by

\begin{equation}
\left(d_{ij}^{*},\mu_{ij}^{*}\right)=\arg\max_{\left(d_{ij},\mu_{ij}\right)}P_{ij}^{\beta\rightarrow+\infty}\left(d_{ij},\mu_{ij}\right)\quad\forall\,\left(i,j\right)\in E
\end{equation}

Unfortunately the computation of \eqref{eq:multi:marg} is impractical
as it would require the calculation of a sum of an exponential number
of terms. We seek to estimate these marginals via the cavity method
approach. We report here a standard formulation of the cavity equations
and we refer the interested reader to \citep{mezard_information_2009}
for the detailed derivation. At finite $\beta$ the BP equations on
our factor graph are:

\begin{equation}
\begin{cases}
m_{ij}\left(d_{ij},\mu_{ij}\right) & =\frac{1}{Z_{ij}}\sum_{\substack{\left\{ d_{ki},\mu_{ki}\right\} :\\
k\in\partial i\setminus j
}
}\psi_{i}\left(\boldsymbol{d}_{i},\boldsymbol{\mu}_{i}\right)e^{-\beta\sum_{\mu}c_{i}^{\mu}\prod_{k\in\partial i}\left(1-\delta_{\mu_{ki},\mu}\right)}\prod_{k\in\partial i\setminus j}n_{ki}\left(d_{ki},\mu_{ki}\right)\\
\\
n_{ki}\left(d_{ki},\mu_{ki}\right) & =e^{-\beta w_{ki}\mathbb{I}\left[d_{ki}>0\right]}m_{ki}\left(d_{ki},\mu_{ki}\right)
\end{cases}\label{eq:bp1}
\end{equation}
where

\[
Z_{ij}=\sum_{\{d_{ij},\mu_{ij}\}}m_{ij}\left(d_{ij},\mu_{ij}\right)
\]
is the normalization constant or ``partial'' partition function.
The functions $m_{ij}$ are called\textit{ cavity marginals} or ``messages'',
suggesting that some information is flowing on edge $\left(i,j\right)$
within the factor graph from node $i$ to node $j$. In fact, the
values of the messages $m_{ij}$ are in some sense proportional to
the probability of a particular assignment $\left(d_{ij},\mu_{ij}\right)$
for edge $\left(i,j\right)$ if the node $j$ were temporarily erased
from the graph.

The system of equations in \eqref{eq:bp1} can be seen as fixed point
equations that can be solved iteratively. Starting from a set of initial
cavity marginals at time $t=0$, we iterate the right-hand-side of
\eqref{eq:bp1} until numerical convergence to a fixed point is reached.
At convergence we calculate an approximation to marginals in \eqref{eq:multi:marg}
via the \textit{cavity fields} defined as

\begin{equation}
M_{ij}\left(d_{ij},\mu_{ij}\right)\propto n_{ij}\left(d_{ij},\mu_{ij}\right)n_{ji}\left(-d_{ij},\mu_{ij}\right)\label{eq:exactmarg}
\end{equation}
where the proportional sign denotes that a normalization constant
is missing. 

Cavity equations for optimization problems can be easily obtained
by substituting the $m_{ij}$ and $M_{ij}$ with the variables $h_{ij}\left(d_{ij},\mu_{ij}\right)=\lim_{\beta\rightarrow\infty}\frac{1}{\beta}\log n\left(d_{ij},\mu_{ij}\right)$
and $H_{ij}\left(d_{ij},\mu_{ij}\right)=\lim_{\beta\rightarrow\infty}\frac{1}{\beta}\log M_{ij}\left(d_{ij},\mu_{ij}\right)$
into \eqref{eq:bp1} and \eqref{eq:exactmarg} that play the role
of cavity marginals and fields in the zero-temperature limit; the
resulting closed set of equations is known as the Max-Sum algorithm.
At convergence we can extract our optimal assignment of variables
by the computation of the\textit{ decisional} variables 

\begin{eqnarray}
\left(d_{ij}^{*},\mu_{ij}^{*}\right) & = & \arg\max_{\left(d_{ij},\mu_{ij}\right)}H_{ij}\left(d_{ij},\mu_{ij}\right)\label{eq:decisional}\\
H_{ij}\left(d_{ij},\mu_{ij}\right) & = & h_{ij}\left(d_{ij},\mu_{ij}\right)+h_{ji}\left(-d_{ij},\mu_{ij}\right)-C'\label{eq:MSmarg}
\end{eqnarray}
where $C'$ is an additive constant that guarantees that normalization
condition in the zero-temperature limit, i.e. $\max_{\left(d_{ij},\mu_{ij}\right)}H_{ij}\left(d_{ij},\mu_{ij}\right)=0$,
is satisfied. In practice, converge is reached when the \textit{decisional}
variables computed as in \eqref{eq:decisional} do not change after
a predefined number of successive iterations (often $10-30$). Notice
that taking the $\beta\rightarrow+\infty$ limit of the message-passing
equations at finite $\beta$ is not equivalent to the zero-temperature
limit of the Boltzmann distribution in \eqref{eq:multi:prob}. 

In the following section we will show how to derive equations for
the cavity marginals and cavity fields, for finite $\beta$ and in
the limit $\beta\rightarrow+\infty$, depending on we are dealing
with the V-DStP or the E-DStP problem. 

\section{The cavity equations \label{sec:Derivation}}

\subsection{Vertex-disjoint Steiner trees Problem \label{subcavity-V-DStP}}

To derive the Belief Propagation equations for the V-DStP problem
suffices to impose $\psi_{i}\left(\boldsymbol{d}_{i},\boldsymbol{\mu}_{i}\right)=\psi_{i}^{V}\left(\boldsymbol{d}_{i},\boldsymbol{\mu}_{i}\right)$
in \eqref{eq:bp1}. By a change of variables, we will determine a
Max-Sum algorithm for this variant.

Equations for messages can be easily obtained by using the properties
of Kronecker delta functions in $\psi_{i}^{V}\left(\boldsymbol{d}_{i},\boldsymbol{\mu}_{i}\right)$;
the explicit derivation in reported in appendix \ref{sec:AppmesspasVertex}.
We can differentiate three cases depending on we are updating messages
$n_{ij}$ for positive, negative or null depth $d_{ij}$:

\begin{equation}
\begin{cases}
m_{ij}\left(d,\mu\right)=m_{+}^{b}\left(d,\mu\right)+m^{f}\left(d,\mu\right) & \quad\forall d>0,\mu\neq0\\
m_{ij}\left(d,\mu\right)=m_{-}^{b}\left(d,\mu\right)+m^{f}\left(d,\mu\right) & \quad\forall d<0,\mu\neq0\\
m_{ij}\left(0,0\right)=e^{-\beta\sum_{\mu}c_{i}^{\mu}}\prod_{k\in\partial i\backslash j}n_{ki}\left(0,0\right)+m_{0}^{b}+m_{0}^{f}
\end{cases}\label{eq:vdbp}
\end{equation}
where $m_{+}^{b}\left(d,\mu\right),\,m_{-}^{b}\left(d,\mu\right),$
$m^{f}\left(d,\mu\right),$ and $m_{0}^{b},\,m_{0}^{f}$ are defined
as

\begin{eqnarray*}
m_{+}^{b}\left(d,\mu\right) & = & \prod_{k\in\partial i\backslash j}\left[n_{ki}\left(d+1,\mu\right)+n_{ki}(0,0)\right]\\
m^{f}\left(d,\mu\right) & = & \delta_{c_{i}^{\mu},0}\sum_{k\in\partial i\backslash j}n_{ki}\left(d,\mu\right)\prod_{l\in\partial i\backslash\left\{ j,k\right\} }n_{li}\left(0,0\right)\\
m_{-}^{b}\left(d,\mu\right) & = & \sum_{k\in\partial i\backslash j}n_{ki}\left(d+1,\mu\right)\prod_{l\in\partial i\backslash\{j,k\}}\left[n_{li}\left(d,\mu\right)+n_{li}(0,0)\right]\\
m_{0}^{b} & = & \sum_{\mu\neq0}\sum_{d<0}m_{-}^{b}\left(d,\mu\right)\\
m_{0}^{f} & = & \sum_{\mu\neq0}\sum_{d<0}\sum_{k\in\partial i\backslash j}n_{ki}\left(d,\mu\right)\sum_{l\in\partial i\backslash\left\{ j,k\right\} }n_{li}\left(-d,\mu\right)\prod_{m\in\partial i\backslash\left\{ k,l,j\right\} }n_{mi}\left(0,0\right)
\end{eqnarray*}

Replacing $h_{ij}(d_{ij},\mu_{ij})=\lim_{\beta\rightarrow+\infty}n_{ij}\left(d_{ij},\mu_{ij}\right)$
in \eqref{eq:vdbp} we obtain the Max-Sum equations:

\begin{equation}
\begin{cases}
h_{ij}\left(d,\mu\right)=\max\left\{ h_{+}^{b}\left(d,\mu\right),h_{+}^{f}\left(d,\mu\right)\right\}  & \quad\forall d>0,\mu\neq0\\
h_{ij}\left(d,\mu\right)=\max\left\{ h_{-}^{b}\left(d,\mu\right),h_{-}^{f}\left(d,\mu\right)\right\}  & \quad\forall d<0,\mu\neq0\\
h_{ij}\left(0,0\right)=\max\left\{ -\sum_{\mu}c_{i}^{\mu}+\sum_{k\in\partial i\backslash j}h_{ki}\left(0,0\right),\;h_{0}^{b},\,\,h_{0}^{f}\right\} 
\end{cases}
\end{equation}
for

\begin{eqnarray*}
h_{+}^{b}\left(d,\mu\right) & = & -w_{ij}+\sum_{k\in\partial i\backslash j}\max\left\{ h_{ki}\left(d+1,\mu\right),h_{ki}\left(0,0\right)\right\} \\
h_{+}^{f}\left(d,\mu\right) & = & -w_{ij}+\log\delta_{c_{i}^{\mu},0}+\max_{k\in\partial i\backslash j}\left\{ h_{ki}\left(d,\mu\right)+\sum_{l\in\partial i\backslash j}h_{li}\left(0,0\right)\right\} \\
h_{-}^{b}\left(d,\mu\right) & = & \max_{k\in\partial i\backslash j}\left[h_{ki}\left(d+1,\mu\right)-w_{ik}+\sum_{l\in\partial i\backslash\{j,k\}}\max\left\{ h_{li}\left(d,\mu\right),h_{li}\left(0,0\right)\right\} \right]\\
h_{-}^{f}\left(d,\mu\right) & = & \log\delta_{c_{i}^{\mu},0}+\max_{k\in\partial i\backslash j}\left\{ h_{ki}\left(d,\mu\right)-w_{ik}+\sum_{l\in\partial i\backslash j}h_{li}\left(0,0\right)\right\} \\
h_{0}^{b} & = & \max_{\mu\neq0}\max_{d<0}h_{-}^{b}\left(d,\mu\right)\\
h_{0}^{f} & = & \max_{\mu\neq0}\max_{d<0}\left[\max_{\substack{k\in\partial i\setminus j\,,\\
l\in\partial i\backslash\{j,k\}
}
}h_{ki}\left(d,\mu\right)-w_{ik}+h_{li}\left(-d,\mu\right)+\sum_{m\in\partial i\backslash\left\{ j,k,l\right\} }h_{mi}\left(0,0\right)\right]
\end{eqnarray*}

\subsection{Edge-disjoint Steiner trees problem \label{subsec:cavity-E-DStP}}

As for the V-DStP, the Belief Propagation equations for the E-DStP
can be computed imposing $\psi_{i}\left(\boldsymbol{d}_{i},\boldsymbol{\mu}_{i}\right)=\psi_{i}^{E}\left(\boldsymbol{d}_{i},\boldsymbol{\mu}_{i}\right)$
into \eqref{eq:bp1}:

\begin{equation}
m_{ij}\left(d_{ij},\mu_{ij}\right)\propto\sum_{\substack{\left\{ d_{ki},\mu_{ki}\right\} :\\
k\in\partial i\setminus j
}
}\psi_{i}^{E}\left(\boldsymbol{d}_{i},\boldsymbol{\mu}_{i}\right)e^{-\beta\sum_{\mu}c_{i}^{\mu}\prod_{k\in\partial i}\left(1-\delta_{\mu_{ki},\mu}\right)}\prod_{k\in\partial i\setminus j}n_{ki}\left(d_{ki},\mu_{ki}\right)\label{eq:BP-packing}
\end{equation}

Instead of considering the cavity messages as in \ref{subcavity-V-DStP},
to compute \eqref{eq:BP-packing} we will first define a partial partition
function

\begin{equation}
Z_{i}=\sum_{\boldsymbol{d}_{i},\boldsymbol{\mu}_{i}}\psi_{i}^{E}\left(\boldsymbol{d}_{i},\boldsymbol{\mu}_{i}\right)e^{-\beta\sum_{\mu}c_{i}^{\mu}\prod_{k\in\partial i}\left(1-\delta_{\mu_{ki},\mu}\right)}\prod_{k\in\partial i}n_{ki}\left(d_{ki},\mu_{ki}\right)\label{eq:Zeta_i}
\end{equation}
and then calculate the set of messages $m_{ij}\left(d,\mu\right)$
(for all possible values of $d$ and $\mu$) from $i$ to $j$ through
\eqref{eq:Zeta_i} by temporarily setting the message that $i$ received
from $j$ as $n_{ji}\left(d_{ji},\mu_{ji}\right)=\delta_{-d,d_{ji}}\delta_{\mu,\mu_{ji}}$.
In fact, here $m_{ij}$ has a unique non-zero value in the state that
satisfies the anti-symmetric property, namely $d_{ij}=d$, $\mu_{ij}=\mu$;
under this condition $Z_{i}=m_{ij}\left(d_{ij},\mu_{ij}\right)$ up
to a normalization constant. Due to the explicit expression of $\psi_{i}^{E}$
message-passing equations become intractable and, therefore, the update
step of the algorithm cannot be efficiently implemented. In the following
subsections we overcome this issue by proposing two different approaches
for the computation of \eqref{eq:Zeta_i} where we make use of two
different sets of auxiliary variables. The first formalism relies
on ``binary occupation'' variables that denote, for each node of
the factor graph, if edges incident on it are used or not by any communication;
as we will see the associated computation scales exponentially in
the degree of the nodes. The second one consists in a mapping between
the E-DStP update equation and a weighted matching problem over bipartite
graphs, that, in the $\beta\rightarrow+\infty$, becomes a weighted
maximum matching problem which can be solved efficiently. This implementation
scales exponentially with respect to $M$ but it may be more efficient
for vertices with large degrees with respect to the first algorithm.

\subsubsection{Neighbors occupation formalism}

Suppose of associating with each vertex $i$ a vector $\boldsymbol{x}=\left\{ 0,1\right\} {}^{|\partial i|}$
that denotes if edges incident on $i$ are employed or unemployed
within the solution. A feasible assignment of these auxiliary variables
is guaranteed if, for every link $\left(i,k\right)\in E$ incident
on $i$, we impose $x_{k}=1$ if the edge belongs to a tree (i.e.
$d_{ki}\neq0$ and consequently $\mu_{ki}\neq0$) or $x_{k}=0$ otherwise
(for $\mu_{ki}=0,\,d_{ki}=0$). Variables $\left(\boldsymbol{d}_{i},\boldsymbol{\mu}_{i}\right)$
must locally satisfy the following identity $\prod_{k\in\partial i}\mathbb{I}\left[x_{k}=1-\delta_{d_{ki},0}\right]=1$
for every node $i\in V$. If we insert this expression in \eqref{eq:Zeta_i}
and we sum over all possible assignments of $\boldsymbol{x}$ variables
we obtain
\begin{eqnarray}
Z_{i} & = & \sum_{\boldsymbol{d}_{i},\boldsymbol{\mu}_{i}}\psi_{i}^{E}\left(\boldsymbol{d}_{i},\boldsymbol{\mu}_{i}\right)e^{-\beta\sum_{\mu}c_{i}^{\mu}\prod_{k\in\partial i}\left(1-\delta_{\mu_{ki},\mu}\right)}\sum_{\boldsymbol{x}}\prod_{j\in\partial i}\mathbb{I}\left[x_{j}=1-\delta_{d_{ji},0}\right]n_{ji}\left(d_{ji},\mu_{ji}\right)\label{eq:Zeta_i_x}\\
 & = & \sum_{\boldsymbol{x}}Z_{\mathbf{x}}^{M}
\end{eqnarray}
where $Z_{\boldsymbol{x}}^{M}$ is defined by computing the following
expression for $q=M$
\begin{equation}
Z_{\boldsymbol{x}}^{q}\equiv\sum_{\substack{\boldsymbol{d}_{i},\boldsymbol{\mu}_{i}\\
\mu_{ki}\leq q
}
}\psi_{i}^{E}\left(\boldsymbol{d}_{i},\boldsymbol{\mu}_{i}\right)e^{-\beta\sum_{\mu}c_{i}^{\mu}\prod_{k\in\partial i}\left(1-\delta_{\mu_{ki},\mu}\right)}\prod_{k\in\partial i}\mathbb{I}\left[x_{k}=1-\delta_{d_{ki},0}\right]n_{ki}\left(d_{ki},\mu_{ki}\right)\label{eq:Zeta_M}
\end{equation}
The computation of $Z_{\boldsymbol{x}}^{q}$ is then performed using
the following recursion (the equivalence of \eqref{eq:Zeta_q} to
\eqref{eq:Zeta_M} is proven in appendix \ref{sec:RecursiveZ})

\begin{eqnarray}
Z_{\boldsymbol{x}}^{q} & = & \sum_{\boldsymbol{y\leq\boldsymbol{x}}}\left(g_{\boldsymbol{y}}^{0}+g_{\boldsymbol{y}}^{b}+g_{\boldsymbol{y}}^{f}\right)Z_{\boldsymbol{y}}^{q-1}\label{eq:Zeta_q}\\
Z_{\boldsymbol{x}}^{0} & = & e^{-\beta\sum_{\mu}c_{i}^{\mu}}\prod_{j\in\partial i}\delta_{x_{j},0}n_{ji}\left(0,0\right)
\end{eqnarray}
where the auxiliary functions $g_{\boldsymbol{y}}^{0},\,g_{\boldsymbol{y}}^{b},\,g_{\boldsymbol{y}}^{f}$
are defined as
\begin{eqnarray*}
g_{\boldsymbol{y}}^{0} & = & e^{-\beta c_{i}^{q}}\prod_{\substack{k\in\partial i\\
y_{k}=0\\
x_{k}=1
}
}n_{ki}\left(0,0\right)\\
g_{\boldsymbol{y}}^{b} & = & \sum_{d>0}\sum_{\substack{j\in\partial i\\
y_{j}=0\\
x_{j}=1
}
}n_{ji}\left(-d,q\right)\prod_{\substack{k\in\partial i\setminus j\\
y_{k}=0\\
x_{k}=1
}
}\left[n_{ki}\left(d+1,q\right)+n_{ki}\left(0,0\right)\right]\\
g_{\boldsymbol{y}}^{f} & = & \delta_{c_{i}^{q},0}\sum_{d>0}\sum_{\substack{j\in\partial i\\
y_{j}=0\\
x_{j}=1
}
}n_{ji}\left(-d,q\right)\sum_{\substack{k\in\partial i\backslash j\\
y_{k}=0\\
x_{k}=1
}
}n_{ki}\left(d,q\right)\prod_{\substack{l\in\partial i\setminus\left\{ j,k\right\} \\
y_{l}=0\\
x_{l}=1
}
}n_{li}\left(0,0\right)
\end{eqnarray*}
and the trace over $\boldsymbol{y}\leq\boldsymbol{x}$ denotes all
possible vectors $\boldsymbol{y}=\left\{ 0,1\right\} {}^{|\partial i|}$
satisfying

\begin{equation}
y_{k}=\begin{cases}
y_{k}\leq x_{k} & \mathrm{if}\,\mu_{ki}\neq q\\
0 & \mathrm{if}\,\mu_{ki}=q
\end{cases}
\end{equation}
 Within the Max-Sum formalism,we can equivalently define a partial
free entropy $F_{i}=\lim_{\beta\rightarrow+\infty}\frac{1}{\beta}\log Z_{i}$
and express it as function of Max-Sum messages $h_{ij}\left(d_{ij},\mu_{ij}\right)=\lim_{\beta\rightarrow+\infty}\frac{1}{\beta}\log n_{ij}\left(d_{ij},\mu_{ij}\right)$
as

\begin{equation}
F_{i}=\max_{\substack{\boldsymbol{d}_{i},\boldsymbol{\mu}_{i}\\
\psi_{i}^{E}\left(\boldsymbol{d}_{i},\boldsymbol{\mu}_{i}\right)=1
}
}\max_{\boldsymbol{x}}\left[\sum_{k\in\partial i}\log\mathbb{I}\left[x_{k}=1-\delta_{d_{ki},0}\right]+h_{ki}\left(d_{ki},\mu_{ki}\right)-\sum_{\mu}c_{i}^{\mu}\prod_{k\in\partial i}\left(1-\delta_{\mu_{ki},\mu}\right)\right]
\end{equation}
where the function $\sum_{k\in\partial i}\log\mathbb{I}\left[x_{k}=1-\delta_{d_{ki},0}\right]$
takes value zero if variables satisfy the constraints or minus infinity
otherwise. As in the BP formulation, we rewrite it as

\begin{equation}
F_{i}=\max_{\boldsymbol{x}}F_{\boldsymbol{x}}^{M}
\end{equation}
with
\[
F_{\boldsymbol{x}}^{M}=\max_{\substack{\boldsymbol{d}_{i},\boldsymbol{\mu}_{i}\\
\psi_{i}^{E}\left(\boldsymbol{d}_{i},\boldsymbol{\mu}_{i}\right)=1
}
}\sum_{k\in\partial i}\left[\log\mathbb{I}\left[x_{k}=1-\delta_{d_{ki},0}\right]+h_{ki}\left(d_{ki},\mu_{ki}\right)-\sum_{\mu}c_{i}^{\mu}\prod_{k\in\partial i}\left(1-\delta_{\mu_{ki},\mu}\right)\right]
\]

The computation can be performed recursively from
\begin{align}
F_{\boldsymbol{x}}^{q} & =\max_{\boldsymbol{y\leq\boldsymbol{x}}}\left\{ F_{\boldsymbol{y}}^{q-1}+\max\left\{ h_{0},\,h_{b},\,h_{f}\right\} \right\} \\
F_{\mathbf{x}}^{0} & =-\sum_{\mu}c_{i}^{\mu}+\log\mathbb{I}\left[\boldsymbol{x}=\boldsymbol{0}\right]+\sum_{k\in\partial i}h_{ki}\left(0,0\right)
\end{align}
where

\begin{align}
h_{0} & =\sum_{\substack{k\in\partial i\\
y_{k}=0\\
x_{k}=1
}
}h_{ki}\left(0,0\right)-c_{i}^{q}\\
h_{b} & =\max_{d>0}\max_{\substack{k\in\partial i\\
y_{k}=0\\
x_{k}=1
}
}\left[h_{ki}\left(-d,q\right)+\sum_{\substack{l\in\partial i\setminus k\\
y_{l}=0\\
x_{l}=1
}
}\max\left[h_{li}\left(d+1,q\right),h_{li}\left(0,0\right)\right]\right]\\
h_{f} & =\log\delta_{c_{i}^{q},0}+\max_{d>0}\left[\max_{\substack{k\in\partial i,\,l\in\partial i,\,k\neq l\\
y_{k}=0,\,y_{l}=0\\
x_{k}=1\,x_{l}=1
}
}h_{ki}\left(-d,q\right)+h_{li}\left(d,q\right)+\sum_{\substack{m\in\partial i\backslash\{k,l\}\\
y_{m}=0\\
x_{m}=1
}
}h_{mi}\left(0,0\right)\right]
\end{align}

\subsubsection{Mapping into a matching problem}

We will develop an alternative method for the computation of the
messages of BP and MS update equations, that can lead to an exponential
speedup in some cases. Let us introduce an auxiliary vector $\boldsymbol{s}\in\left\{ 0,1,\dots,D\right\} ^{M}$
associated with each vertex of the graph. Components $s_{\mu}$ take
value in the set of the possible positive depths $\left\{ 1,\ldots,D\right\} $
if this node is member of communication $\mu$ or $0$ otherwise.
For a node $i$ that is not a root but a member of the communication
$\mu$, there exists exactly one neighbor $k$ such that $d_{ik}>0$,
$d_{ki}=-s_{\mu_{ki}}\,\mu_{ki}=\mu$ and for the remaining ones,
$d_{li}\delta_{\mu_{li},\mu}=s_{\mu_{ki}}+1$ or $d_{li}\delta_{\mu_{li},\mu}=0$,
$l\in\partial i\backslash k$. The compatibility function for E-DStP
can be expressed as a function of the new variables as
\begin{eqnarray}
\psi_{i}^{E}\left(\boldsymbol{d}_{i},\boldsymbol{\mu}_{i}\right) & = & \prod_{\mu=1}^{M}\left[\sum_{s_{\mu}>0}\sum_{k\in\partial i}\delta_{\tilde{d}_{ki},-s_{\mu}}\prod_{l\in\partial i\setminus k}\left(\delta_{\tilde{d}_{li},s_{\mu}+1}+\delta_{\tilde{d}_{li},0}\right)+\prod_{k\in\partial i}\delta_{\tilde{d}_{ki},0}\right]\\
 & = & \sum_{\mathbf{s}}\left\{ \prod_{\mu=1}^{M}\left(1-\delta_{s_{\mu},0}\right)\sum_{k\in\partial i}\left[\delta_{\tilde{d}_{ki},-s_{\mu}}\prod_{l\in\partial i\setminus k}\left(\delta_{\tilde{d}_{li},s_{\mu}+1}+\delta_{\tilde{d}_{li},0}\right)\right]+\prod_{\mu=1}^{M}\delta_{s_{\mu},0}\prod_{k\in\partial i}\delta_{\tilde{d}_{ki},0}\right\} \label{eq:Psi-Ed-match}
\end{eqnarray}

As introduced in the neighbors occupation formalism, we will compute
the update equations from $Z_{i}$, that, within this formalism can
be written as

\begin{eqnarray}
Z_{i} & = & \sum_{\boldsymbol{s}}R_{\boldsymbol{s}}Z_{\boldsymbol{s}}\label{eq:Zeta_i_s}
\end{eqnarray}
where 
\begin{eqnarray}
R_{\boldsymbol{s}} & = & \prod_{k\in\partial i}\left[\sum_{\nu}n_{ki}\left(s_{\nu}+1,\nu\right)+n_{ki}\left(0,0\right)\right]\\
Z_{\boldsymbol{s}} & = & \sum_{\boldsymbol{t}}\prod_{\mu}e^{-\beta c_{i}^{\mu}\mathbb{I}\left[s_{\mu}=0\right]}\mathbb{I}\left[\sum_{k\in\partial i}t_{k\mu}=1-\delta_{s_{\mu},0}\right]\prod_{k\in\partial i}\mathbb{I}\left[\sum_{\mu}t_{k\mu}\leq1\right]\prod_{k\in\partial i}\left[\frac{n_{ki}\left(-s_{\mu},\mu\right)}{\sum_{\nu}n_{ki}\left(s_{\nu}+1,\nu\right)+n_{ki}\left(0,0\right)}\right]^{t_{k\mu}}
\end{eqnarray}

The components $t_{k\mu}$ of vectors $\boldsymbol{t}$ take value
1 if $k\in\partial i$ participate to sub-graph $\mu$ at distance
$s_{\mu}>0$ from the root or 0 otherwise. The derivation of \eqref{eq:Zeta_i_s}
is reported in appendix \ref{sec:From-E-DStP-to}. The term $Z_{\boldsymbol{s}}$
is the partition function of a matching problem on the complete bipartite
graph $G=\left(V=A\cup B,E=A\times B\right)$ with $A=\partial i$
and $B=\left\{ \mu:s_{\mu}>0\right\} $, and the energy of a matching
is
\[
\epsilon\left(\boldsymbol{t}\right)=\sum_{k\mu}t_{k\mu}\log\frac{n_{ki}\left(-s_{\mu},\mu\right)}{\sum_{\nu}n_{ki}\left(s_{\nu}+1,\mu\right)+n_{ki}\left(0,0\right)}-\beta c_{i}^{\mu}\mathbb{I}\left[s_{\mu}=0\right]
\]

Notice that the partition function $Z_{\mathbf{s}}$ is computationally
intractable as it corresponds to the calculation of a matrix \emph{permanent.
}In the $\beta\to\infty$ limit we can introduce the Max-Sum messages
$h_{ki}\left(-s_{\mu},\mu\right)=\frac{1}{\beta}\log n_{ki}\left(-s_{\mu},\mu\right)$
and directly compute the free entropy $F_{i}=\frac{1}{\beta}\log Z_{i}$
that fortunately reduces to the evaluation of
\begin{eqnarray}
F_{i} & = & \max_{\boldsymbol{s}}\left[\frac{1}{\beta}\left(\log R_{\boldsymbol{s}}+\log Z_{\boldsymbol{s}}\right)\right]\\
 & = & \max_{\mathbf{s}}\left\{ \sum_{k\in\partial i}\max\left[\max_{\mu}h_{ki}\left(s_{\mu}+1,\mu\right),\;h_{ki}\left(0,0\right)\right]+F_{\mathbf{s}}\right\} 
\end{eqnarray}
The second term $F_{\boldsymbol{s}}=\frac{1}{\beta}\log Z_{\boldsymbol{s}}$
is the free entropy of the solution of a weighted maximum matching
problem on a bipartite graph which can be performed in polynomial
time (precisely, in $O\left(\left(M+\left|\partial i\right|\right)^{2}M\left|\partial i\right|\right))$.
Indeed, for each assignment of the $\boldsymbol{s}$ we can define
the weights $w_{k\mu}$ associated with each edge $(k,\mu)$ as

\begin{equation}
w_{k\mu}=\begin{cases}
h_{ki}\left(-s_{\mu},\mu\right)-\max_{\nu}\max\left\{ h_{ki}\left(s_{\nu}+1,\nu\right),h_{ki}\left(0,0\right)\right\}  & \mbox{ if }s_{\mu}>0\\
-c_{i}^{\mu} & \mbox{ if }s_{\mu}=0
\end{cases}\label{eq:match:w-1}
\end{equation}
and solve 
\begin{equation}
\begin{cases}
F_{\boldsymbol{s}}=\max\sum_{\left(k,\mu\right)}w_{k\mu}t_{k\mu} & :\\
\sum_{k\in\partial i}t_{k\mu}\leq1 & \quad\forall\mu\\
\sum_{\mu}t_{k\mu}\leq1 & \quad\forall k\in\partial i
\end{cases}\label{eq:int-lin-match}
\end{equation}

The system in \eqref{eq:int-lin-match} describes an integer linear
problem for the resolution of a bipartite maximum weighted matching
problem . Its relaxation to real variables $\boldsymbol{t}$can be
efficiently solved and moreover the optimal solution is proven to
be integer, that is for binary $\boldsymbol{t}$.

\subsection{The parameter $D$}

The \textit{branching} formalism introduced in \ref{sec:Representation}
relies on a parameter $D$ that denotes the maximum allowed distance
between the root and the leaves of any tree. This parameter limits
the depth of solution-trees and therefore the goodness of the results:
a small value for $D$ may prevent the connection of some terminals
but a large value of $D$ will slow down the algorithm affecting the
converge. Thus the value of $D$ needs to be carefully designed to
ensure good performances. Although there is not a clear technique
able to predict the best setting, some heuristics have been proposed
in recent works to determine a minimum feasible value of $D$ for
the MStP and PCStP \citep{biazzo_performance_2012}. In this work,
we adopt methods described in \citep{braunstein_practical_2016} to
find a minimum value of $D_{\mu}$ for each communication $\mu$ and
we than set $D=\max_{\mu}D_{\mu}$. 

It is clear that the computing cost of both V-DStP and E-DStP strongly
depends on the value of $D$, more precisely linearly for the V-DStP
and the binary occupation formalism and polynomially for the matching
problem formulation for the E-DStP, and it could be still prohibit
for graph with large diameter. Fortunately, the use of the\textit{
flat} formalism allows us to reduce the parameter $D$ to $D=\max_{\mu}\left|T_{\mu}\right|$
being $\left|T_{\mu}\right|$ the number of terminals of communication
$\mu$. A proof of this property is reported in \citep{braunstein_practical_2016}
for the single tree problem.

\section{Max-Sum for loopy graphs \label{sec:Max-Sum-for-loopy}}

The goodness of the approximation of the marginals is strictly related
to the properties of the factor graph over which we run the Belief
Propagation algorithm. BP is exact on tree graphs but nevertheless
benefits from nice convergence properties even on general, loopy,
graphs that are locally tree-like \citep{weiss_optimality_2001}.
In the framework of the PCStP and multiple trees variants, there are
several instances of practical interest, such as square or cubic lattices
(2D or 3D graphs) modelling VLSI circuits, where many very short loops
exist and the assumption of negligible correlation among variables
is not satisfied. In many of these cases MS fails to converge in most
of the trials or it requires a prohibitive run-time \citep{braunstein_practical_2016}. 

We employ here a \textit{reinforcement} scheme \citep{bayati_statistical_2008,biazzo_performance_2012}
that is able to make the algorithm converge on a tunable amount of
time with the drawback that the solution may be sub-optimal in terms
of cost. From the viewpoint of the factor graph it adds an extra factor
to edge-variables that acts as an external field oriented in the direction
of the cavity fields of past iterations. It slightly modifies the
original problem into an easier one where a feasible assignment of
variables is more likely to occur. The strength of this perturbation
increases linearly in time in a way that, after few iterations, first
inaccurate predictions will be neglected but, after many iterations
of MS, it let the algorithm converge to, hopefully, an assignments
of variables satisfying all the constraints. We report in section \ref{subsec:Reinforcement}
how to modify the Max-Sum equations for the V-DStP and E-DStP for
including the reinforcement factor.

The reinforcement or bootstrapping procedure described in the following
sub-section is generally sufficient to guarantee convergence on random
networks. In practice, however, MS did not converge in some benchmark
instances, even adopting the bootstrapping procedure. In \citep{braunstein_practical_2016}
we have shown how to complement the MS equations with heuristics to
solve PCStP instances in an efficient and competitive way. At each
iteration we perform a re-weight of node prizes and edge weights according
to temporarily Max-Sum predictions and we then apply a heuristics
to find a tree connecting all nodes of the modified graph. After a
pruning procedure, we obtain a pruned minimum spanning tree which
is surely a feasible candidate solution for the PCStP. The motivation
is based on the fact that although Max-Sum often outputs inconsistent
configurations while trying to reach the optimal assignment of variables,
it still contains some valuable information. Heuristics have the responsibility
of adjusting the assignments of the temporarily decisional variables
guaranteeing a tree-structured solution for any iteration of the main
algorithm. Furthermore heuristics results do not depend on the parameter
$D$ of the model and they can provide solution-trees of any diameter.
We show in section \ref{subsec:Max-Sum-based-heuristics} how to generalize
the combination of Max-Sum and heuristics in the case of multiple
trees for the V-DStP and the E-DStP.

\subsection{Reinforcement \label{subsec:Reinforcement}}

Within the bootstrap procedure and for each iteration $t$ of the
algorithm,we compute the messages $h_{ij}^{t}$ and the cavity fields
$H_{ij}^{t}$ as functions of the original messages $\bar{h}_{ij}^{t}$
as

\begin{align}
h_{ij}^{t}\left(d_{ji},\mu_{ji}\right) & \leftarrow\bar{h}_{ji}^{t}\left(d_{ij},\mu_{ji}\right)+\gamma_{t}H_{ji}^{t-1}\left(d_{ji},\mu_{ji}\right)\\
H_{ij}^{t}\left(d_{ij},\mu_{ij}\right) & \propto\bar{h}_{ij}^{t}\left(d_{ij},\mu_{ij}\right)+\bar{h}_{ji}^{t}\left(-d_{ij},\mu_{ij}\right)+\gamma_{t}H_{ij}^{t-1}\left(d_{ij},\mu_{ij}\right)
\end{align}
The parameter $\gamma_{t}=t\gamma_{0}$ is linearly proportional to
$\gamma_{0}$ which is the reinforcement factor that governs the strength
of the bootstrap. It is usually very small, of the order of $10^{-5}$
not to deviate the dynamics towards the minimum of the energy and
thus affect the goodness of the solution. 

\subsection{Max-Sum based heuristics \label{subsec:Max-Sum-based-heuristics} }

At each iteration of the main algorithm we perform a re-weighting
of the graph to favor MS temporarily predictions and we then apply
two fast heuristics to find as many spanning trees as the number of
communications that we want to pack. These trees will be carefully
pruned in order to decrease the cost of the solution. In this work
we design two different re-weighting schemes for two different heuristics
and we refer the interested reader to the single-tree heuristics explained
in \citep{braunstein_practical_2016} for additional details. For
each sub-graph $\mu$ we apply one of the two schemes as follows.

\subsubsection{Shortest Path Tree }

For any MS iteration $t$ we compute the auxiliary weights of a sub-graph
$\mu$ as

\begin{equation}
w_{ij}^{t}=\max_{d\neq0}\left|H_{ij}^{t}\left(d,\mu\right)\right|\label{eq:weightsSPT}
\end{equation}
Notice that since the field $H_{ij}^{t}$ is normalized, there exists
only one assignment of the variables such that $H_{ij}^{t}\left(d^{*},\mu\right)=0$
in correspondence of the most probable state $\left(d^{*},\mu^{*}\right)$;
the field computed in all remaining states will be as negative as
the probability of the corresponding edge to be not employed in the
(temporary) MS solution. For this reason we allow edges in \eqref{eq:weightsSPT}
to have zero weights if they are likely to be exploited within communication
$\mu$ (they would have $d^{*}\neq0$); differently, we penalize edges
that, according to MS, must not be used (for which $d^{*}=0$) imposing
strictly positive weights equal to the (minus) MS field computed at
the most probable non-zero depth, which corresponds, in this case,
to the second maximum of $\max_{d}\left|H_{ij}^{t}\left(d,\mu\right)\right|$.

We then compute the Shortest Paths Tree (SPT) of the modified graph
and we prune the solution tree removing a leaf $i$ if it is not a
terminal (for the MStP), and edges $\left(i,j\right)$ satisfying
$w_{ij}>c_{i}^{\mu}$ (for the PCStP); we repeat this procedure until
we do not find such leaves. 

\subsubsection{Minimum Spanning Tree}

In this scheme we assign auxiliary costs to nodes of the graph according
to MS prediction. Let us consider the two auxiliary functions

\begin{equation}
\begin{cases}
h_{i}\left(d,\mu\right)=\max_{k\in\partial i}\left\{ h_{ik}^{t}(-d,\mu)+\sum_{l\in\partial i\setminus k}\max\left[h_{li}^{t}\left(d+1,\mu\right),h_{li}^{t}\left(0,\mu\right)\right]\right\}  & \quad\mbox{ for }d>0\\
h_{i}\left(0,\mu\right)=\sum_{k\in\partial i}h_{ki}^{t}(0,\mu)-c_{i}^{\mu}
\end{cases}
\end{equation}

A node satisfying $\max_{d>0}h_{i}\left(d,\mu\right)<h_{i}\left(0,\mu\right)$
will be penalized assigning to edges incident on it a large cost $C$.
We then apply the Minimum Spanning Tree (MST) algorithm to the modified
graph and we prune the solution as in the case of the SPT.

Heuristics are applied to the graph for all the communications providing,
for both E-DStP and V-DStP, a superposition of single-tree solutions.
Notice that heuristics are sequentially applied, i.e. we consider
one communication at the time, and depending on we are dealing with
V-DStP or E-DStP, edges (and Steiner nodes for the V-DStP) selected
in the first spanning trees cannot be further used by the successive
applications. To overcome this problem, we add an erasing step before
the application of each heuristics in which we delete edges (and eventually
Steiner nodes) used by other communications. For V-DStP we only need
to cut edges incident on terminals of other sub-graphs to satisfy
nodes-disjoint constraints. Unfortunately such strong edge cutting
procedure may lead to a graph with disconnected components or a graph
in which the terminals that we aim at connecting may be isolated.
In these scenarios we cannot find further trees able to span the modified
graph and thus this heuristic approach fails. One way of preventing
this problem is to randomize the order of the trees over which we
apply the heuristics.

\section{Numerical results}

In this section we report the results for several experiments on synthetic
networks and on benchmark, real-world, instances for the VLSI. In
all the cases we will solve the V-DStP or the E-DStP where terminals
have infinite prizes, i.e. the MStP variant, and a predefined root
is selected for each sub-graph. The synthetic networks we chose are
fully connected, regular or grid graphs, whose properties will allow
us to underline the main features of the models and formalisms introduced
in this work. In particular, by means of the fully connected graphs
we will illustrate the improvements carried by Max-Sum against a ``greedy''
search of the solutions introduced in section \ref{subsec:Greedy-algorithm}
Furthermore, regular graphs allow us to verify the different scaling
of the running time with respect to the degree of the graph of the
two algorithms presented for the E-DStP. Motivated by their importance
on technological applications, namely in the the design of VLSI, we
also show some results on grid, both synthetic and real-word, graphs:
here we will underline the improvements carried by the \textit{flat}
model. Generally, energies are averaged over several instances, meaning
different realizations of the weighting of the edges and assignment
of the terminals, of the same graph. To measure the energy gap of
the solutions found by the two different procedures, for instance
``x'' and ``y'' algorithms, we measure the quantity $\frac{E_{x}-E_{y}}{E_{y}}$
assuming that $E_{x}$ and $E_{y}$ are the energies of the solutions
found by algorithm ``x'' and ``y'' respectively. If the gap is
positive (negative) the ``x'' (``y'') algorithm outperforms the
other one.

We underline that, due to the intrinsic difficulty of the problem,
there are very few (exact or approximate) results in literature and
few algorithms to use for the comparison. In the case of VLSI circuits,
we report the solution costs of a state-of-the art linear programming
technique for the V-DStP published in \citep{hoang_steiner_2012}.
This algorithm is not publicly available and thus it cannot be used
for further comparisons.

\subsection{Greedy algorithm\label{subsec:Greedy-algorithm}}

The ``greedy'' procedure consists in solving, for each communication
of the graph, the corresponding single-tree MStP by means of the MS
algorithm combined with the bootstrap. To ensure that the superposition
of these trees is a feasible candidate solution for the packing problem,
we performed, as in the case of the heuristics described in section \ref{subsec:Max-Sum-based-heuristics},
a pre-processing of the graph before any application of the single-tree
algorithm. In particular, whenever we try to propose a solution for
the sub-graph $\mu$, we cut any terminal node, and all edges incident
on it, of the communications that have not yet been considered, together
with edges (and Steiner tree nodes for the V-DStP variant) of the
communications that we have already connected. The ``greedy'' energy
is given by the sum of energies of single-tree solutions.Notice that
this ``greedy'' procedure is actually as hard as the packing problem,
since even the MStP belongs to NP-hard class of problem; nevertheless
this procedure will be useful to underline the benefits carried by
the parallel (packing) search against the ``greedy'' and sequential
one. 

\subsection{Fully connected graphs \label{subsec:FC_results}}

Here we report results for the V-DStP on fully connected graphs where
we aim at packing $M=3$ trees. We compare our performances against
the ``greedy'' procedure. 

We deal with fully connected graphs because here the existence of
a trivial solution of the packing problem, consisting in a chain of
terminal nodes, is always guaranteed. We perform two different experiments:
we first fix the size of the graphs (500 nodes) and we study how energies
and gaps change for an increasing number of terminals nodes. Secondly,
we fix the fraction of terminals per communication, more precisely
for $\alpha=\frac{T_{\mu}}{N}=0.08,\,\mu\in\left\{ 1,2,3\right\} $
and we compare the performances as we increase the size of the graphs
(from 100 to 700 nodes). We run both algorithms with fixed parameters
$D=\{3,5,10\}$ and fixed reinforcement factor $\gamma_{0}=10^{-5}$. 

\subsubsection{Uncorrelated edge weights }

\begin{figure}[h]
\begin{centering}
\includegraphics[width=0.5\textwidth]{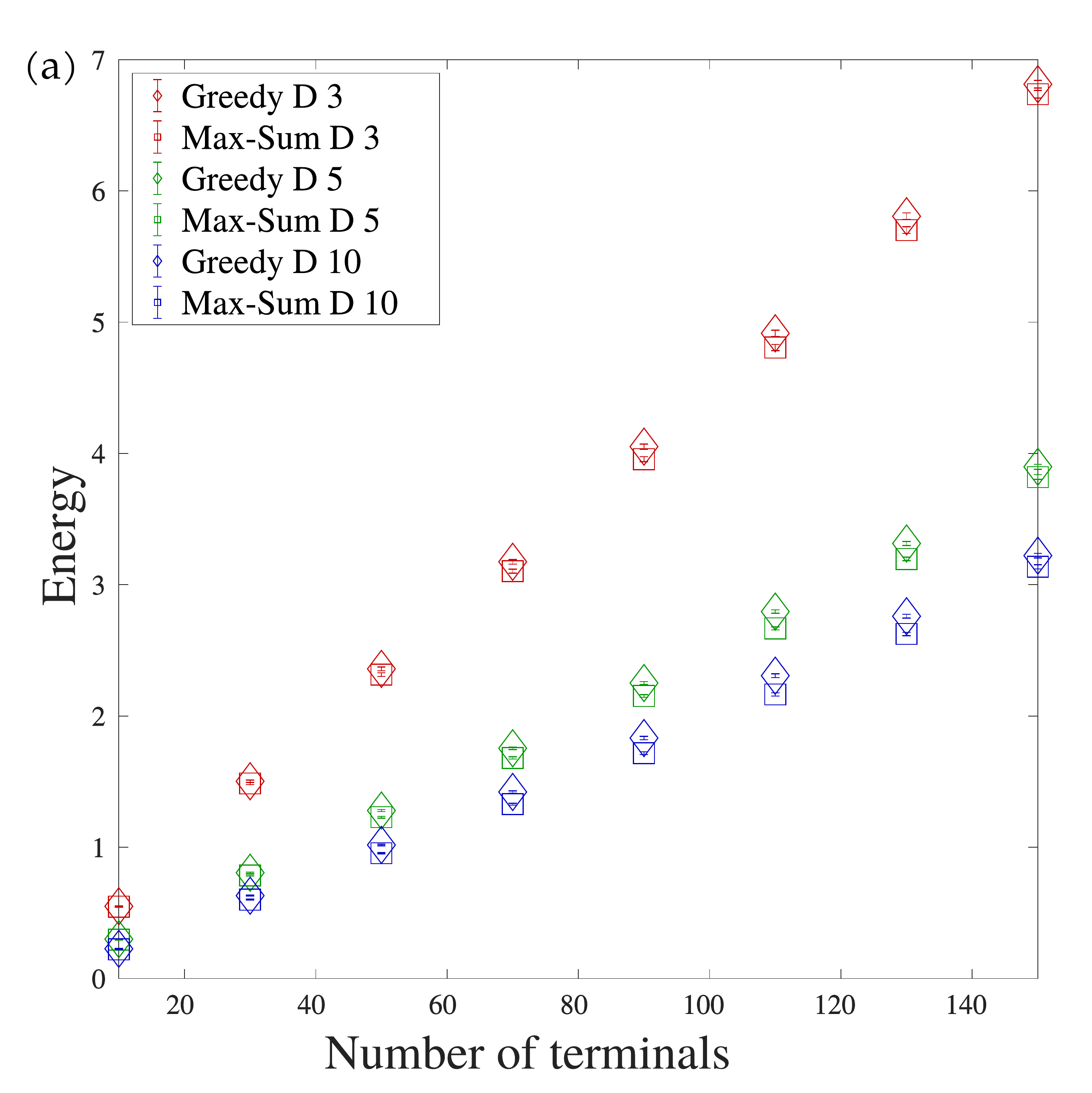}\includegraphics[width=0.49\textwidth]{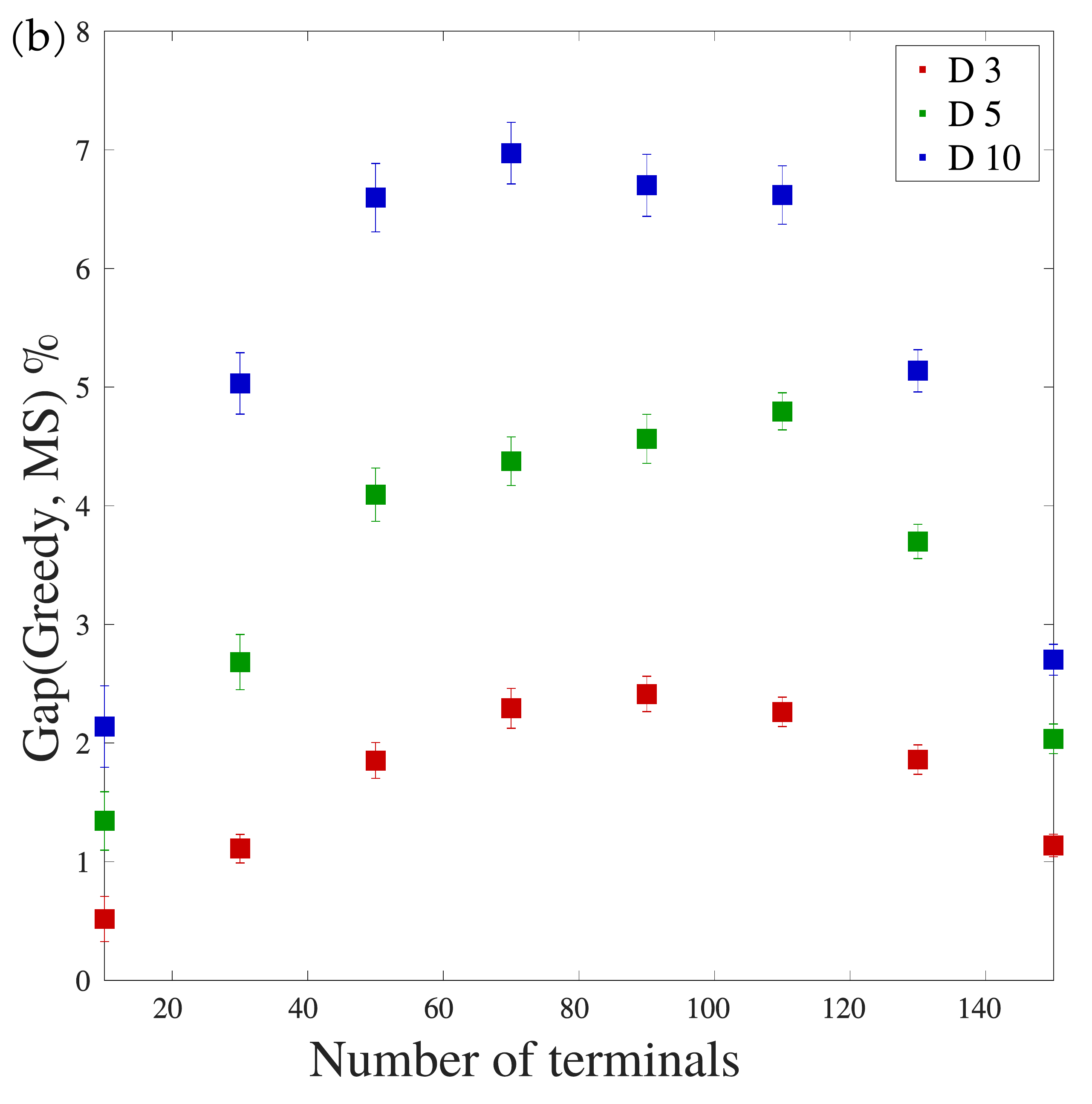}
\par\end{centering}
\caption{Energy (a) and energy gap (b) of the solutions of Max Sum and Greedy
algorithm as functions of the number of terminals. The test instances
are fully connected graphs of 500 nodes with uncorrelated edge weights.
Gaps reported in panel (b) are always positive suggesting that solutions
found by the global search are cheaper in terms of cost than the greedy
ones. \label{fig:En-Gap-FC}}
\end{figure}

These experiments are performed on fully connected graphs where weights
associated with edges are independently and uniformly distributed
random variables in the interval $\left(0,1\right)$. In this scenario,
energies obtained by the greedy procedure are always larger than the
ones achieved by the parallel search, for all values of the number
of terminals and for any value of the parameter $D$ used, as it is
suggested by the plot of the gaps (right plot) in figure \ref{fig:En-Gap-FC}.
Notice that the gaps are slightly greater than zero suggesting that
solutions found by the two methods are very similar in terms of energy
cost, as reported in the plot in figure \ref{fig:En-Gap-FC}, left
panel.

\subsubsection{Correlated edge weights }

\begin{figure}[h]
\begin{centering}
\includegraphics[width=0.51\textwidth]{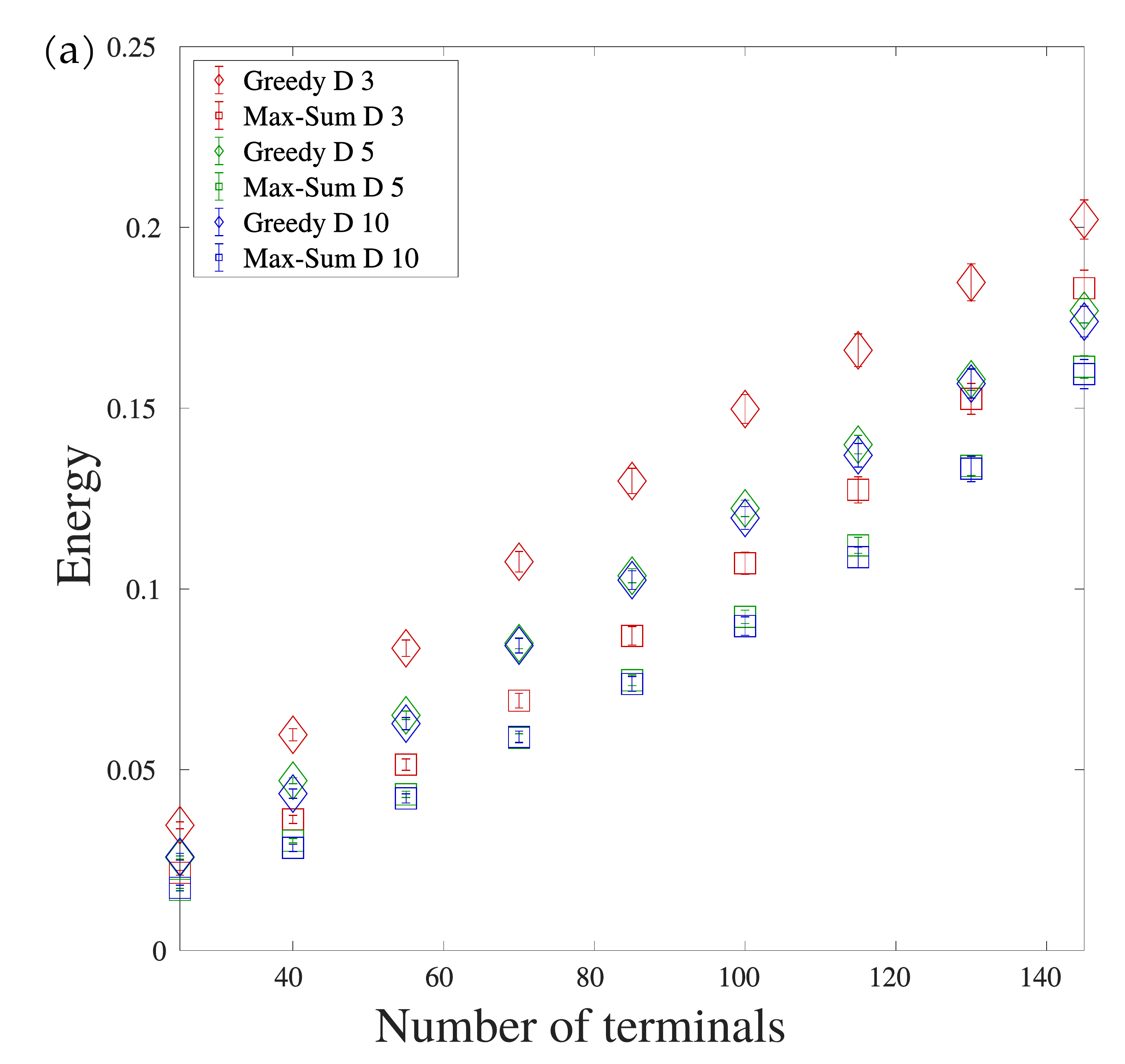}\includegraphics[width=0.5\textwidth]{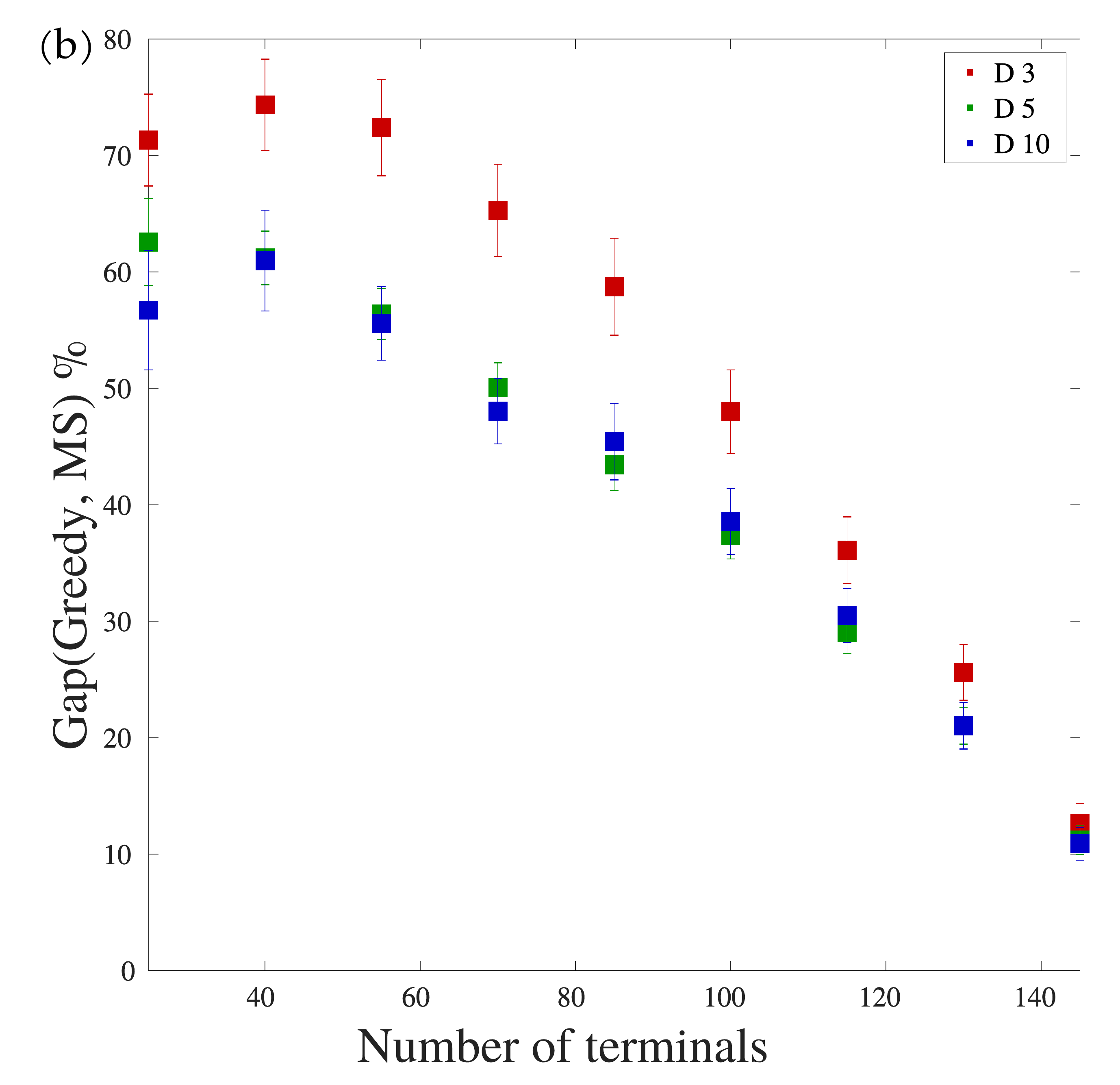}
\par\end{centering}
\caption{Energy (a) and energy gap (b) for Max Sum results against Greedy results
as functions of the number of terminals for correlated edge weighting.
The energy gaps of panel (b) are positive and notably large. \label{fig:EN-FCW}}
\end{figure}

To underline the benefits carried by the optimized strategy, we run
reinforced and greedy reinforced Max-Sum on complete graphs with correlated
edge weights. With each node $i$ we assign a uniformly distributed
random variable $x_{i}$ in the interval $\left(0,1\right)$ and for
each edge $\left(i,j\right)$ we pick a variable $y_{ij}\in\left(0,1\right)$.
Then an edge $\left(i,j\right)$ will be characterized by a weight
$w_{ij}=x_{i}x_{j}y_{ij}$. Here we expect that the cheapest edges
will be chosen by the ``greedy'' algorithm for the solution of the
first trees and, as we proceed with the sequential search, the algorithm
will become the more and more forced to use the remaining expensive
edges. In fact, as shown in figure \ref{fig:EN-FCW}, the gaps notably
increase of one order of magnitude for most of the number of terminals
considered in these experiments. 

Notice that energies encountered for $D=\left\{ 5,10\right\} $ are
very close to one another suggesting that a further increasing of
the parameter $D$, and thus of the solution space, will not lead
to a significant improvement of the solutions.

\subsubsection{Fixed fraction of terminals}

\begin{figure}[h]
\begin{centering}
\includegraphics[width=0.5\textwidth]{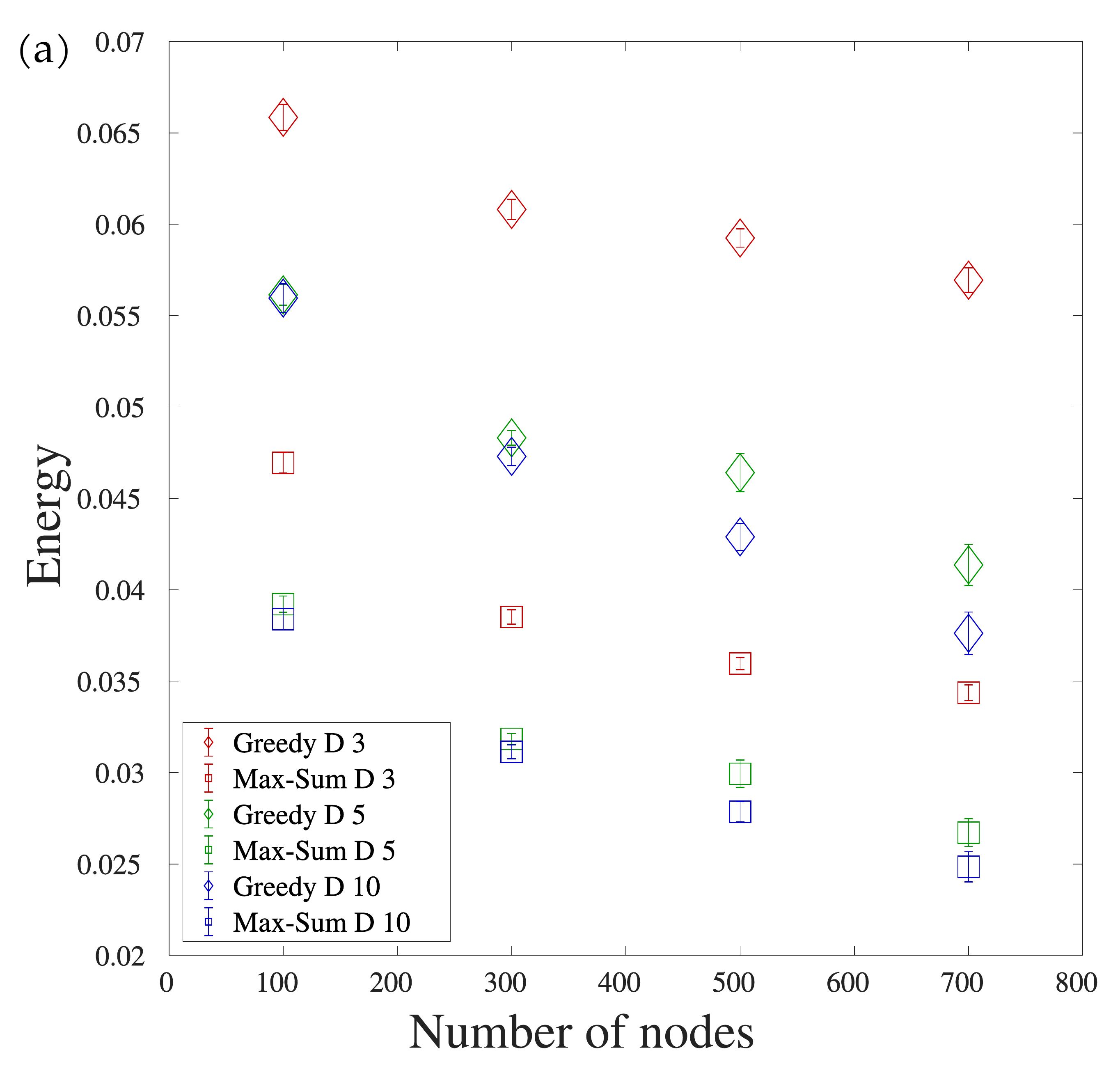}\includegraphics[width=0.5\textwidth]{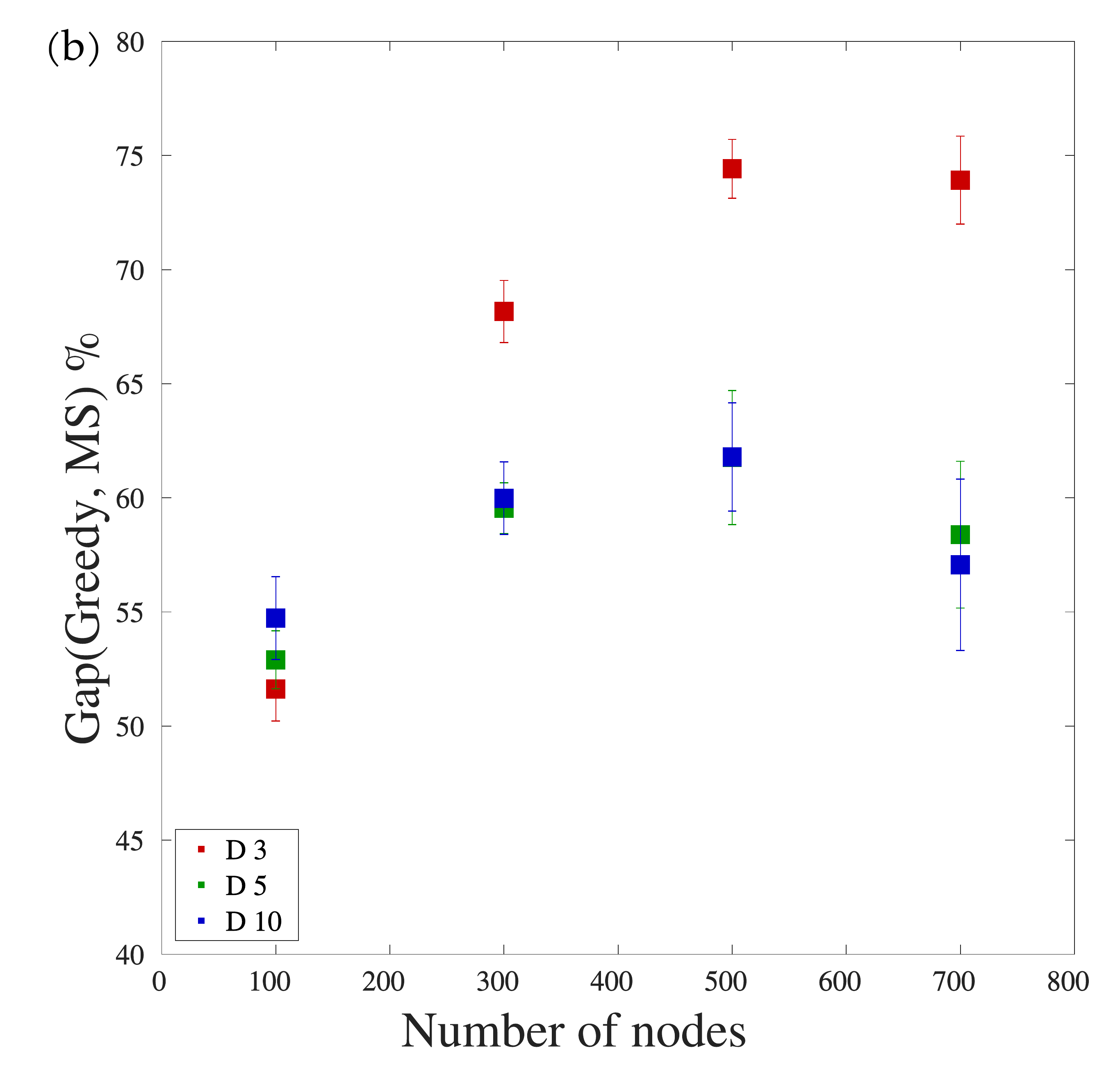} 
\par\end{centering}
\caption{Energy (a) and energy gap (b) for Max Sum against Greedy results as
functions of the number of nodes for a fixed fractions of terminals
(per communication)$\alpha=0.08$. \label{fig:Energy-en-gap-fixed-fraction}}
\end{figure}

To study the performances in the asymptotic limit, namely for $N\rightarrow+\infty$
,$T_{\mu}\rightarrow+\infty$ for each communication $\mu$ and constant
$\alpha$, we attempted the solution of V-DStP on complete graphs
having a fixed fraction of terminals $\alpha=0.08$ and for an increasing
number of nodes $N$. Although non-rigorous, this procedure can suggest
us the behavior of the energies and the energy gaps in the large $N$
limit. As reported in figure \ref{fig:Energy-en-gap-fixed-fraction}
panel (a), when the number of nodes reaches $N\in[500,\,700]$, the
energy of both Max Sum and greedy solutions, for all values of $D$,
seems to stabilize to a constant value. As a consequence, as plotted
in figure \ref{fig:Energy-en-gap-fixed-fraction} panel (b), also energy
gaps fluctuates around a fixed value that seems to be different if
one considers $D=3$ or $D=\left\{ 5,10\right\} $. 

\subsection{Regular graphs}

In section \ref{subsec:cavity-E-DStP} we have seen how to deal with
the update equations of BP and MS algorithms for the E-DStP with the
help of two different auxiliary set of variables. Although the final
expressions of the equations are very different, the energies obtained
by both algorithms must be identical; the only differences rely on
the computational cost that strongly depends on the properties of
the graph, precisely on the degree of the nodes of the graph and on
the number of communications. To underline these two features of the
neighbors occupation formalism and matching problem mapping, that
from now will be denoted as \textit{NeighOcc} and \textit{Matching}
algorithms, we perform two different experiments on regular, fixed
degree, graphs for different values of the degree and of the number
of sub-graphs. For these simulations we have fixed the values of the
parameter $D=10$ and the reinforcement factor $\gamma_{0}=10^{-4}$.

\subsubsection{Energy as a function of the degree}

Similarly to the experiments in section \ref{subsec:Energy-as-a}, here
we consider regular graphs of $N=50$ nodes containing $M=3$ sub-graphs
for four possible degrees $d\in\left\{ 3,4,5,6\right\} $. The energies
provided by \textit{NeighOcc} and \textit{Matching} and plotted in figure
\ref{fig:Reg-E-DStP}, panel (a), can be statistically considered
the same, as for the fixed degree experiment shown before. Here the
computational costs (panel (b) and (c) of figure \ref{fig:Reg-E-DStP})
scales exponentially only for the \textit{NeighOcc }(as it is remarked
by the linear trend in the semi-log plot) while it scales polynomially
for the \textit{Matching} formalism as predicted by the analysis on
the update equations in section \ref{subsec:cavity-E-DStP}.

\subsubsection{Energy as a function of the number of communications \label{subsec:Energy-as-a}}

\begin{figure}[h]
\begin{centering}
\includegraphics[width=1\textwidth]{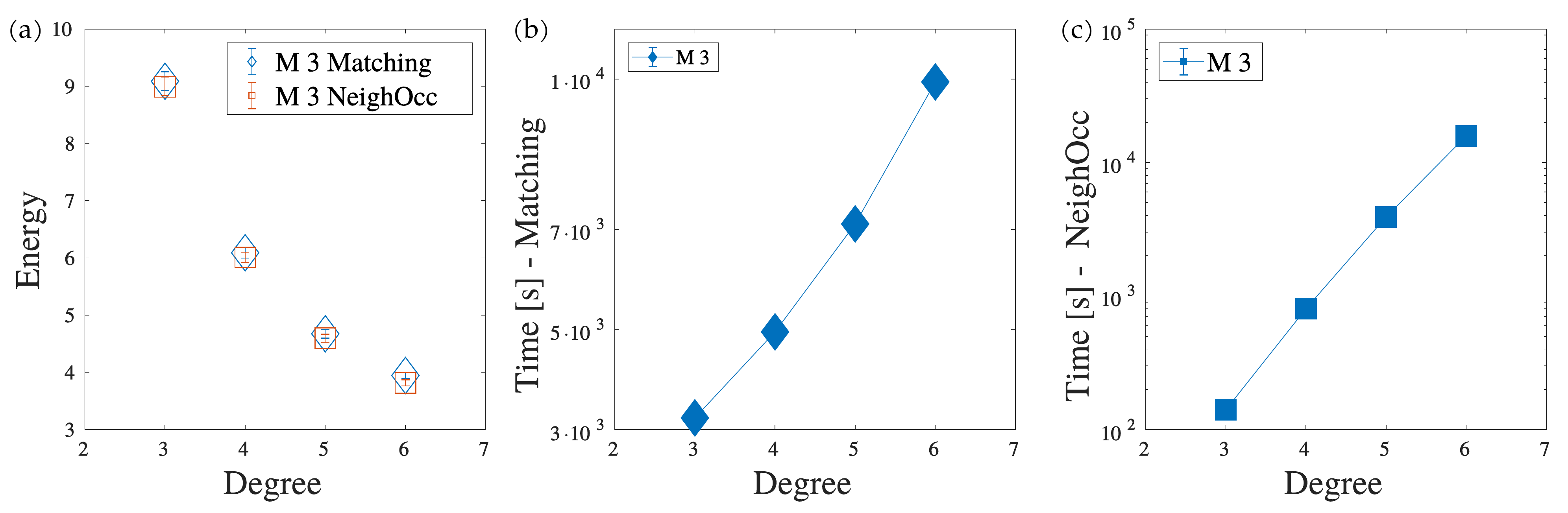}
\par\end{centering}
\begin{centering}
\includegraphics[width=1\textwidth]{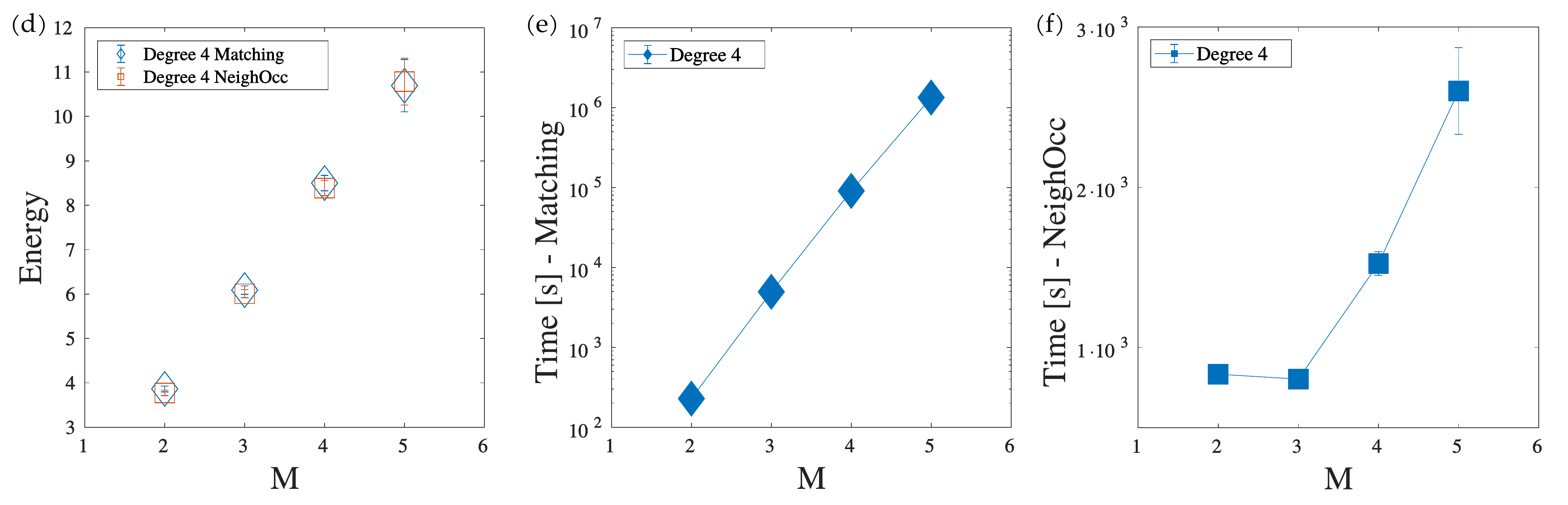}
\par\end{centering}
\caption{In panel (a), energy of the solutions for the E-DStP on regular graphs
of 3 communications as a function of the degree. In (b) and (c) running
time of \textit{Matching} and \textit{NeighOcc} algorithms as a functions
of degree. In panel (d): energy of the solutions for the E-DStP on
regular graphs as a function of the number of packed trees. Panels
(e) and (f): running time of the\textit{ Matching} and \textit{NeighOcc}
algorithms as a functions of the number of communications. In all
plots energies of the solutions are almost the same, but the computing
time dramatically differ as we are using the \textit{Matching} formalism
or the \textit{NeighOcc} algorithm. \label{fig:Reg-E-DStP} }
\end{figure}

In this experiment we try to solve the E-DStP on two sets of regular
graphs of $N=50$ nodes having fixed degree $4$, for an increasing
number of trees. Each communication has the same number of terminals
$T=3$. As shown in figure \ref{fig:Reg-E-DStP}, panel (d), the energy
costs of the solutions provided by \textit{NeighOcc} and \textit{Matching}
algorithms are almost identical as we expected. At the same time,
the computing time plotted in figure \ref{fig:Reg-E-DStP}, panels
(e) and (f), shows that the \textit{Matching }procedure needs a time
that scales exponentially, i.e. linearly in an log-scale plot, on
the number of sub-graphs while it becomes polynomial for the \textit{NeighOcc
}algorithm\textit{. }

\subsection{Grid graphs}

This section is devoted to the illustration of results of both V-DStP
and E-DStP on 2D and 3D lattices. The first experiments are performed
on synthetic 3D lattices of dimension $5\times5\times5$ containing
$N=125$ nodes. Here we fix the number of communications $M\in\left\{ 2,3,4\right\} $
and we study how energies behave when the number of terminals $T$
per communication changes in the range $\left[3,\,6\right]$. For
the V-DStP and the E-DStP (only in the \textit{NeighOcc} formalism)
we compare the results provided by the \textit{branching} and \textit{flat}
models. While the parameter $D$ can be arbitrary large for the \textit{branching}
model, we keep the value $D=T$ for the \textit{flat} one since, as
discussed in section \ref{subsec:The-flat-formalism}, it is sufficient
to explore all the solution space. In the second part of this section
we comment the performances of the MS algorithm and of the MS-based
heuristics presented in section \ref{subsec:Max-Sum-based-heuristics}
applied to several benchmark instances for the design problem of VLSI
circuits. 

\subsubsection{Branching and flat models for the V-DStP e E-DStP (neighbors occupation
formalism)}

\begin{figure}[h]
\centering{}\includegraphics[width=0.49\textwidth]{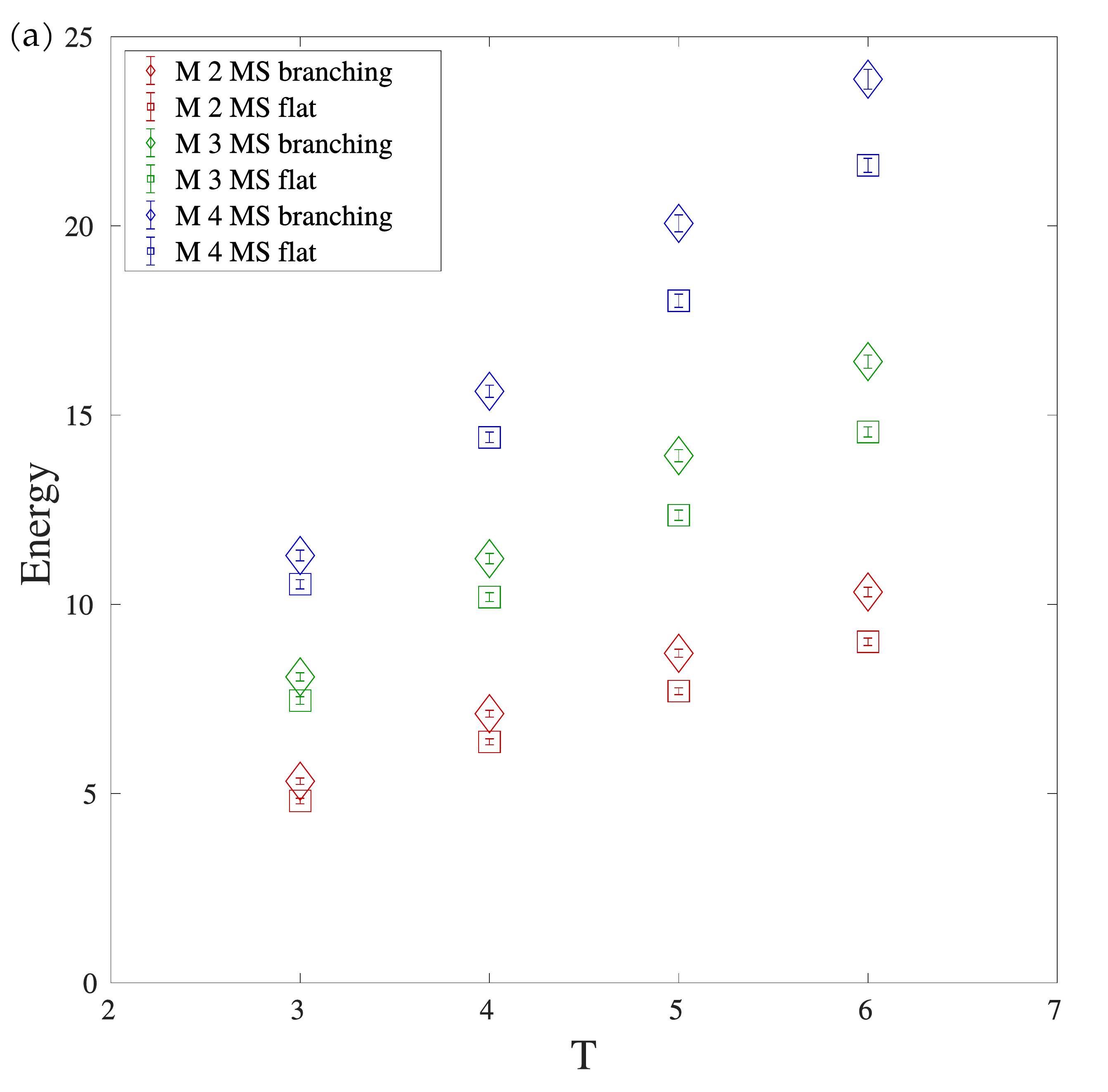}\includegraphics[width=0.5\textwidth]{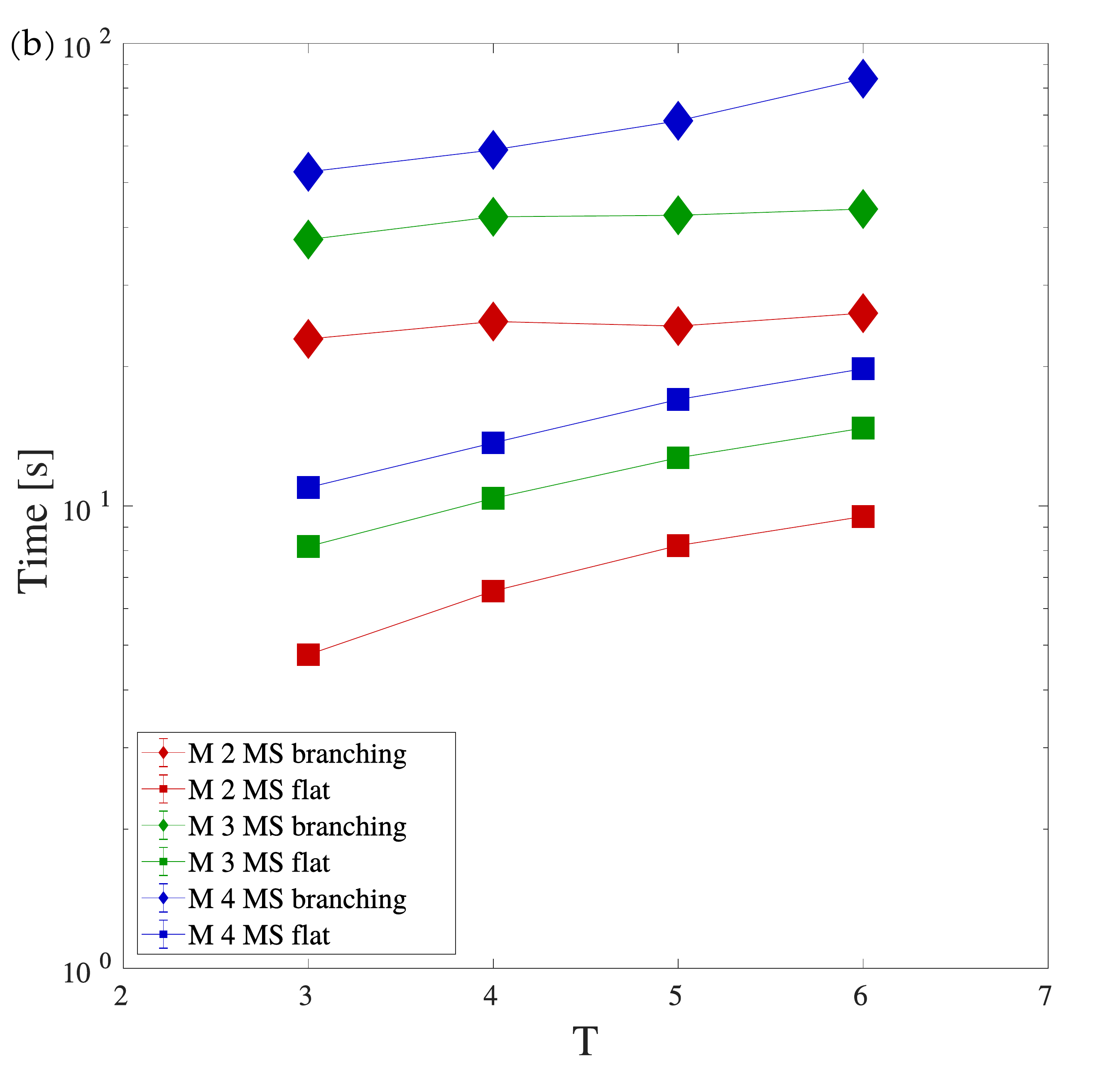}\caption{Energy (a) and computational time (b) as a function of the number
of terminals per communications for 3D grid graphs, V-DStP variant.
\label{fig:V-DStP-branch-flat}}
\end{figure}

As shown in figure \ref{fig:V-DStP-branch-flat}, left panel, the energies
of the solutions found by the \textit{flat} model for the V-DStP are
always smaller than the energies found by the \textit{branching} one.
We underline that, as plotted in the right panel of figure \ref{fig:V-DStP-branch-flat},
the flat version of MS equations has the advantage of converging in
a running time that is always smaller than the one needed by the branching
model. This is reasonable as the parameter $D$, which linearly influences
the computation time of both algorithms, is often greater (on average
$D=8$) for the \textit{branching} model than the one fixed for the
flat representation. 

\begin{figure}[h]
\begin{centering}
\includegraphics[width=0.49\textwidth]{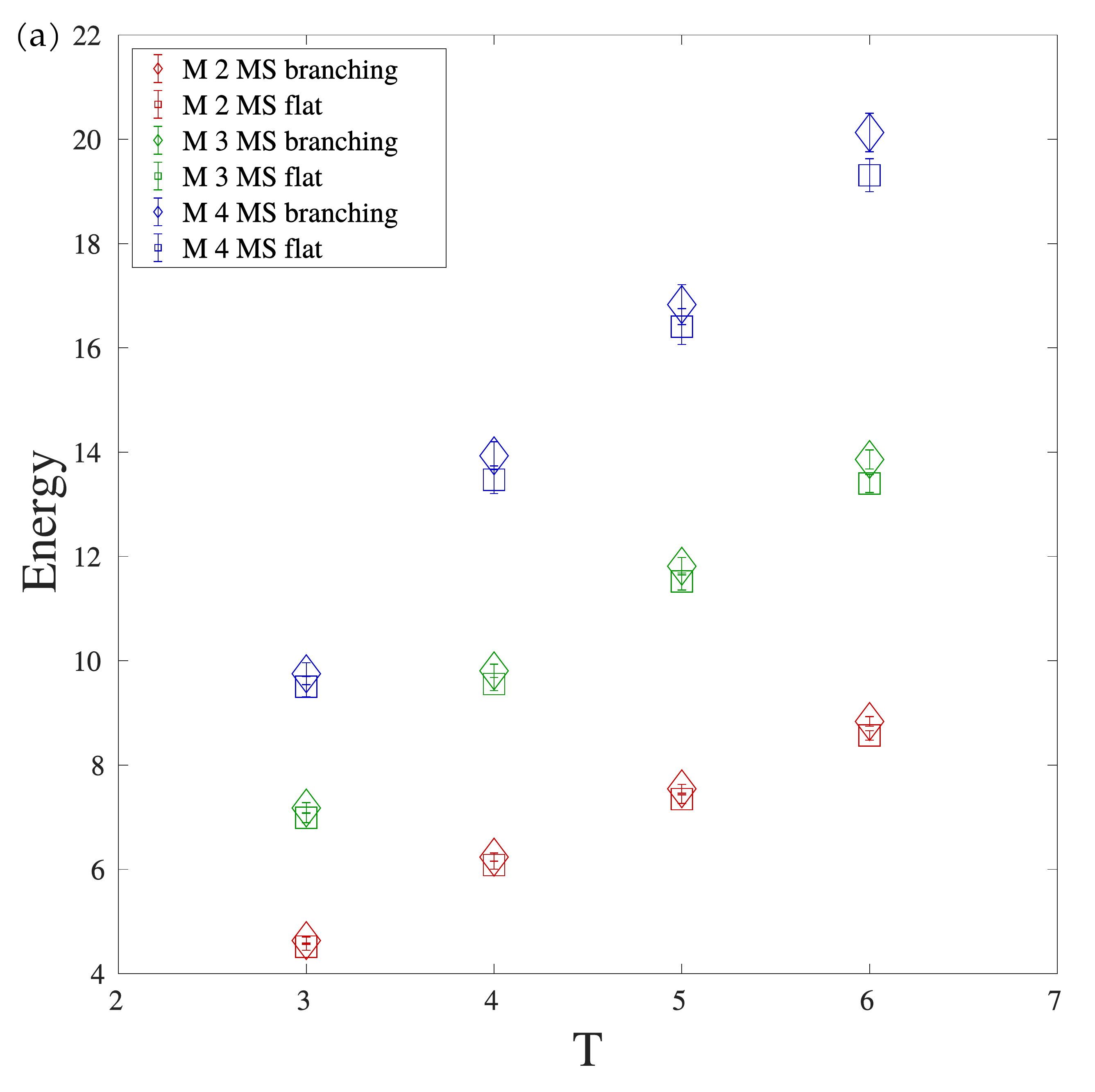}\includegraphics[width=0.5\textwidth]{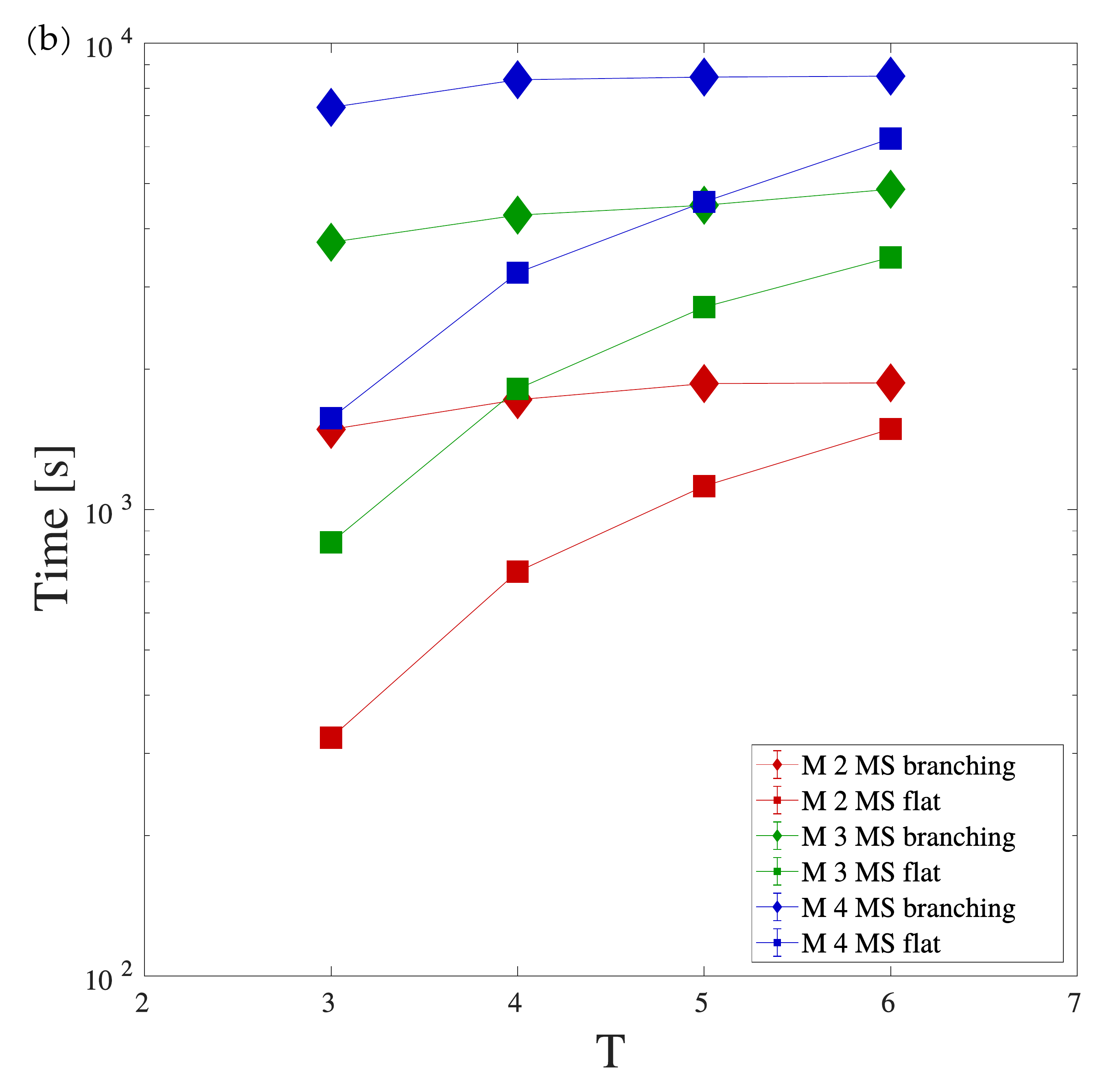}
\par\end{centering}
\caption{Energy (a) and computational time (b) as a function of the number
of terminals per communications for 3D grid graphs, E-DStP variant.
\label{fig:E-DStP-branch-flat}}
\end{figure}

A different behavior is observed for the resolution of the E-DStP
on grids using our two models. As remarked in figure \ref{fig:E-DStP-branch-flat},
left panel, energies found by the flat and branching representations
are comparable; here the depth used by the branching model, on average
equal to $D=8\,\mathrm{and\,}D=9$ for $T\in\left[3,4\right]\,\mathrm{and\,T\in\left[5,6\right]}\,$
respectively, probably suffices to explore the same solution space
considered by the flat formalism for smaller $D$. Still, the flat
model is preferable as it requires a computing time that is smaller
than the one needed by the branching model for all the cases we have
considered.


\subsubsection{V-DStP for VLSI circuits \label{subsec:VLSI_results}}

\begin{figure}[h]
\begin{centering}
\includegraphics[width=0.5\columnwidth]{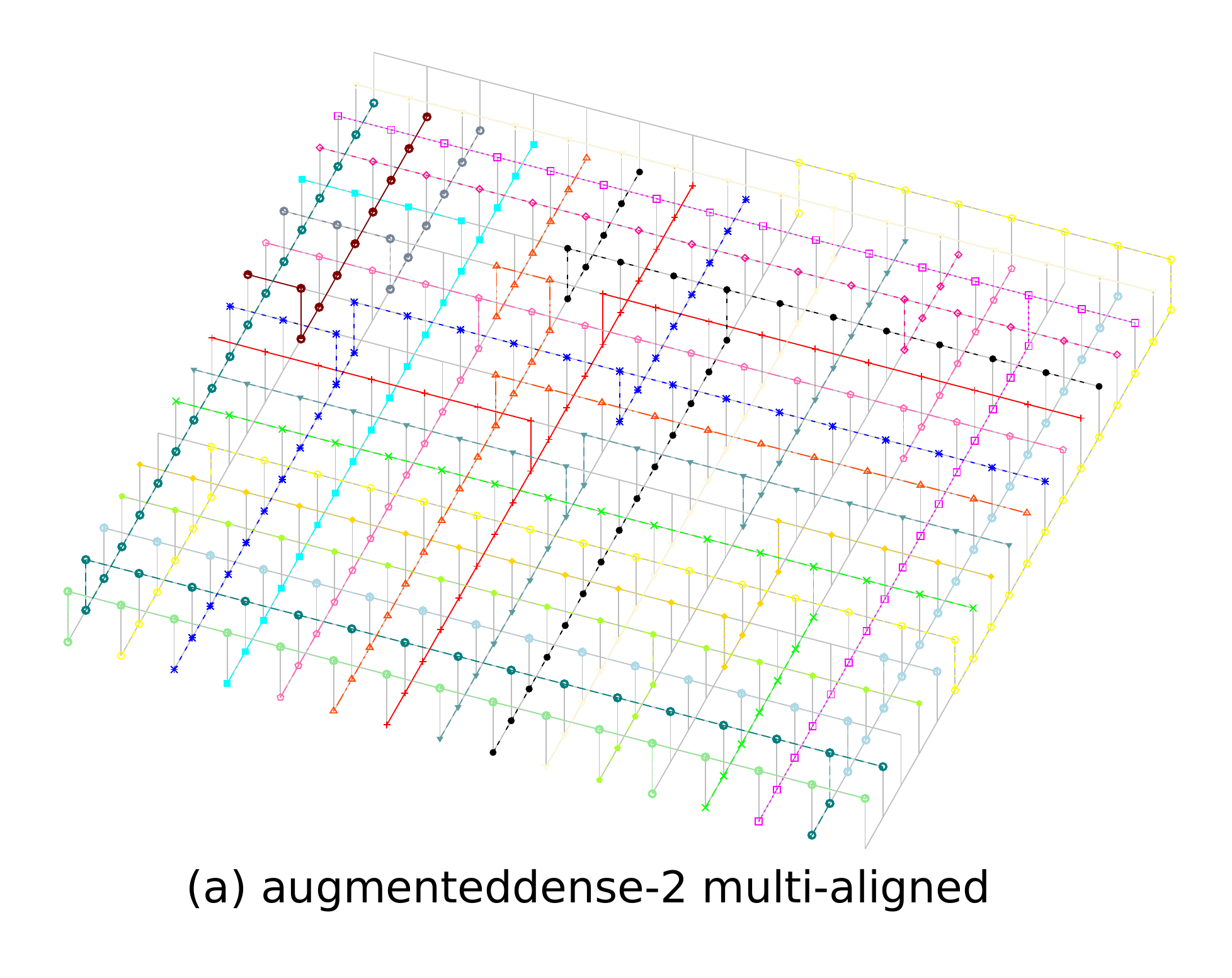}\includegraphics[width=0.5\columnwidth]{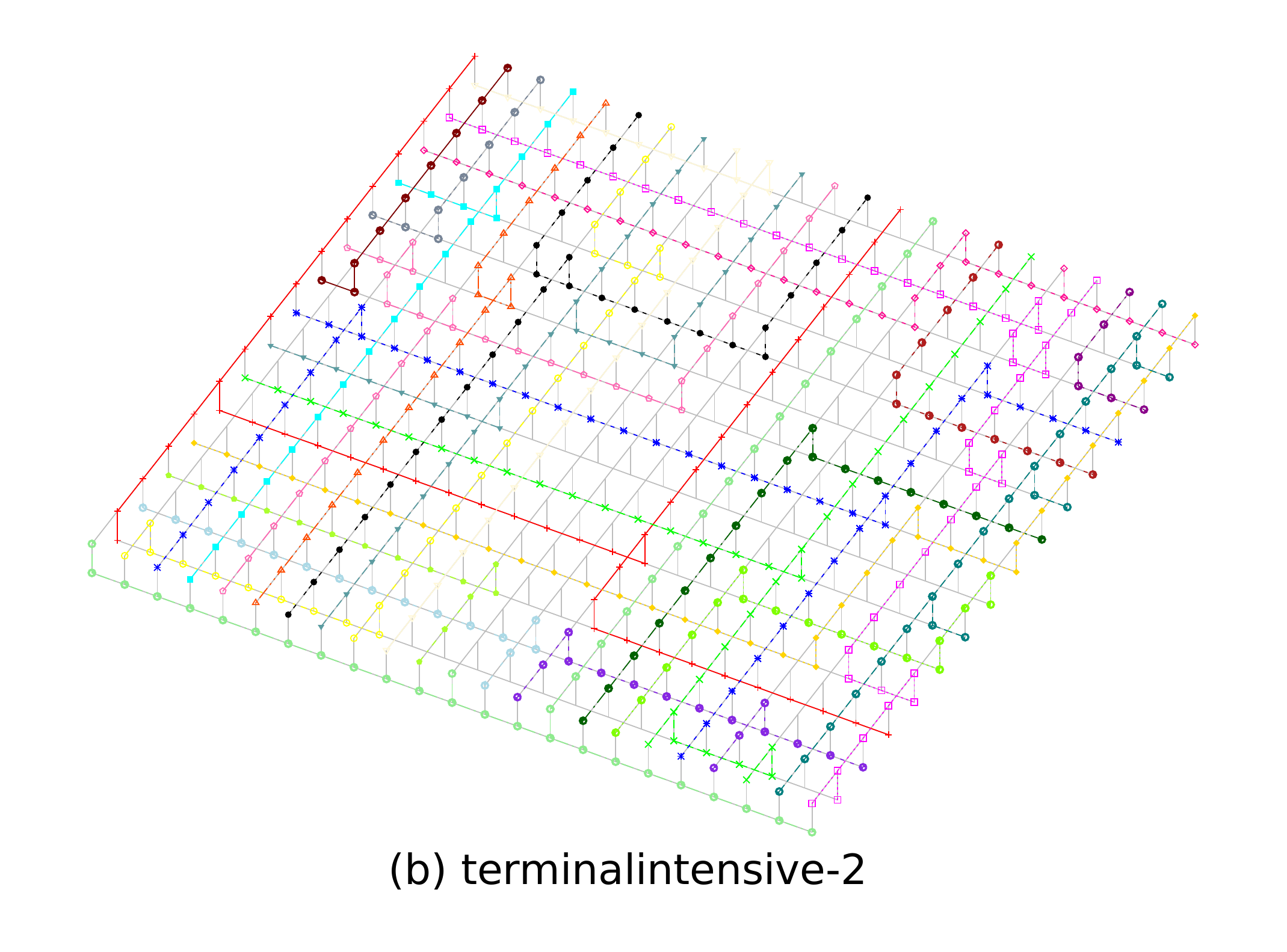}
\par\end{centering}
\begin{centering}
\includegraphics[width=0.5\columnwidth]{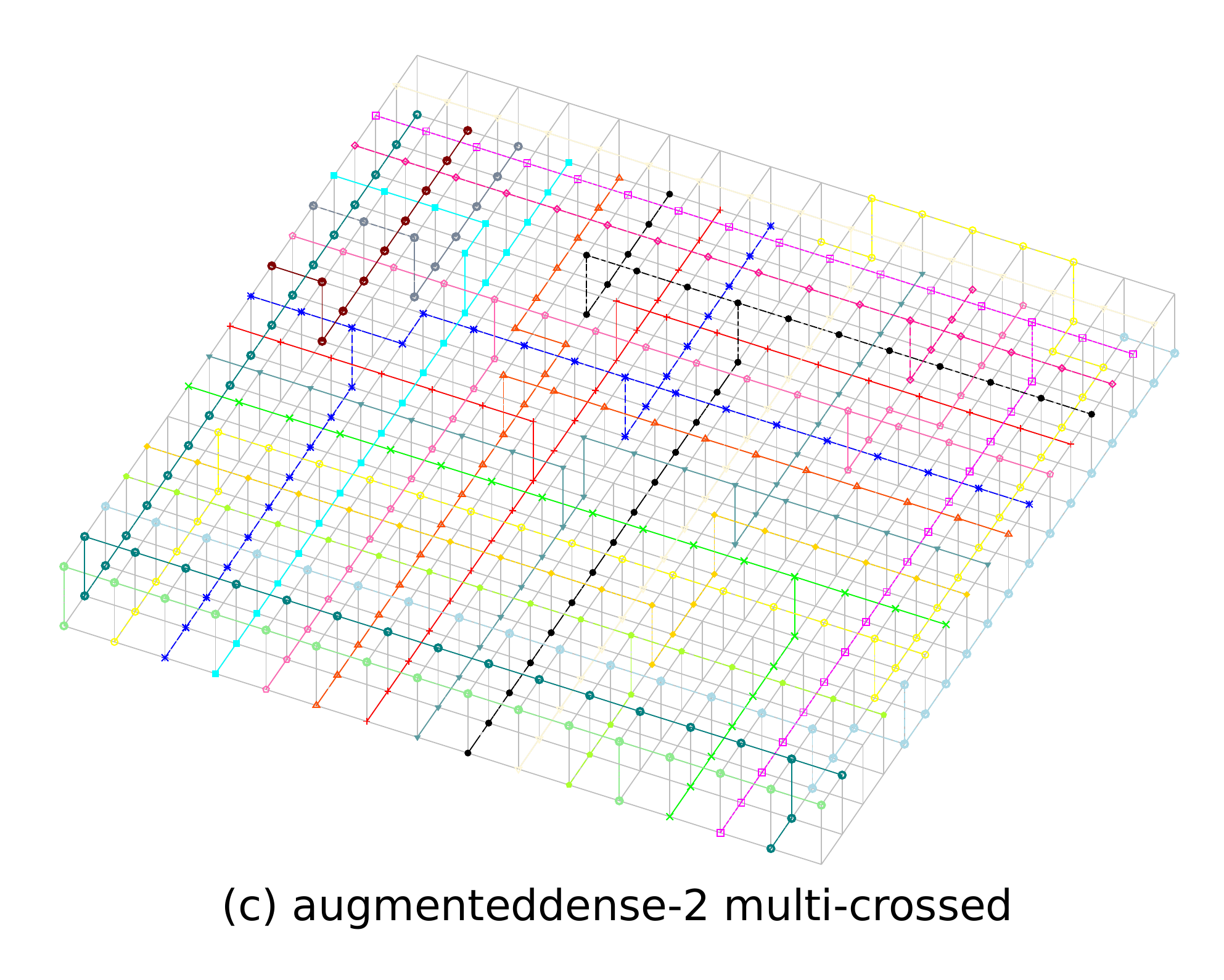}\includegraphics[width=0.5\columnwidth]{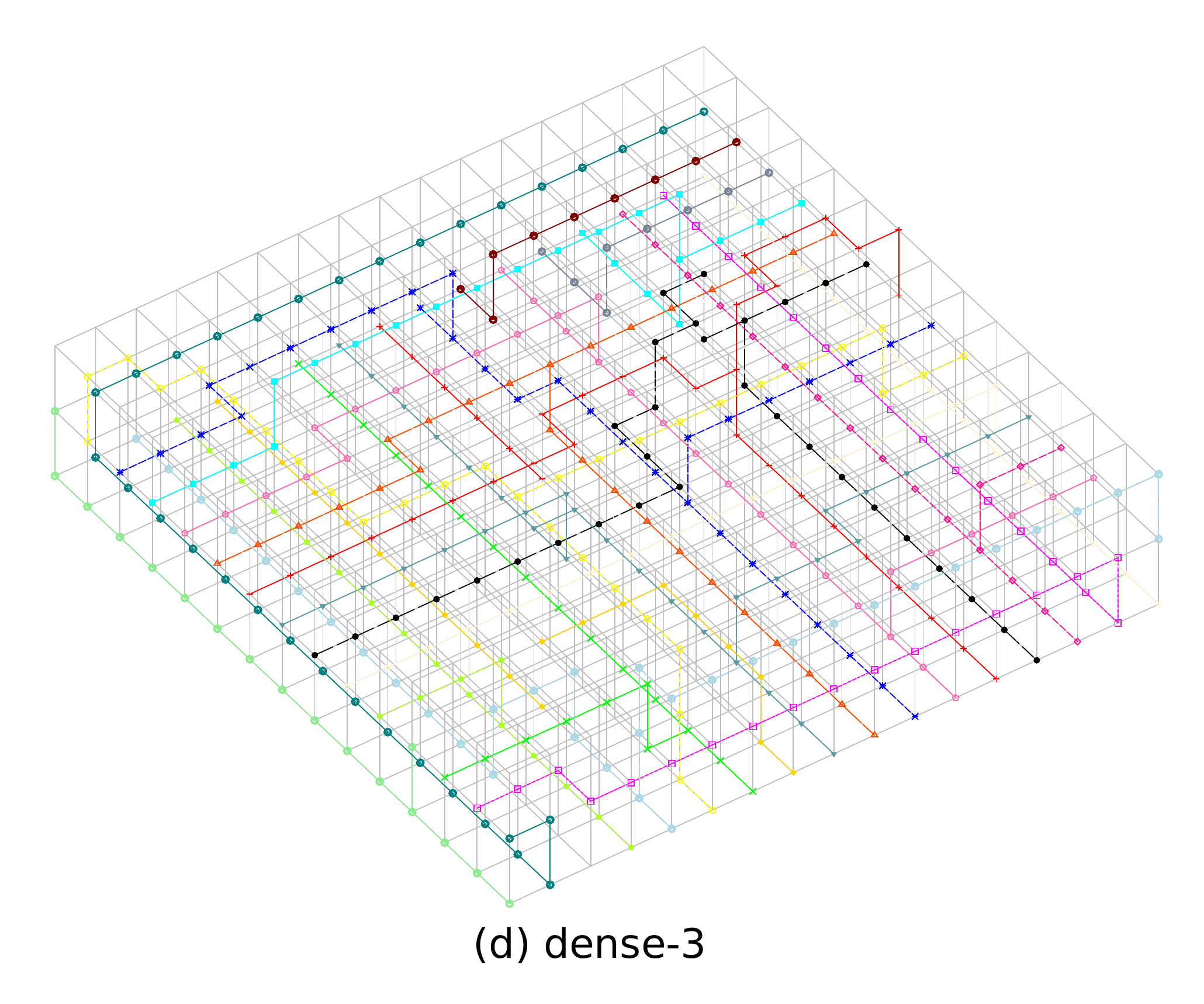}
\par\end{centering}
\caption{Examples of solutions for the V-DStP on VLSI circuits for multi-aligned,
(a) and (b) figures, and multi-crossed, (c) and (d) figures, layouts} \label{grids}
\end{figure}

In this section we report several results for standard benchmark instances
of circuit layout where we solve the V-DStP. Instances are 3D grid
graphs modelling VLSI chips where we pack relatively many trees, usually
19 or 24, each of which typically contains few terminal nodes (3 or
4). Such grid graphs can be seen as multi-layers graphs where we allow
two different kinds of connections. In the \textit{multi-crossed}
layers, each node is connected to all its possible neighbors in all
directions: the resulting graphs are cubic lattices. The \textit{multi-aligned}
layers are similar to the multi-crossed ones but in each layer we
allow only connections in one direction, either east-to-west or north-to-south
\citep{hoang_steiner_2012}. For sake of simplicity, consider a cubic
lattice in a three dimensional Cartesian coordinate system: depending
on the value of the $z-$coordinate, the allowed connections will
be present in directions parallel to the $x$ or to the $y$ axes.
In table \ref{tab:Koch} we first report some information (type of
the layers, size, number of sub-graphs and total number of terminals)
concerning each instance and our results. We show the energies achieved
by reinforced Max Sum along with the ones of the two heuristics described
in section \ref{subsec:Max-Sum-based-heuristics}; in analogy with \citep{braunstein_practical_2016},
we label as ``J'' heuristics that performs a modified SPT and as
``N'' if instead we use the MST. Energies obtained using the \textit{flat}
model are labeled as ``(f)'' while if nothing is specified or ``(b)''
is used, we made use of the \textit{branching} representation. Results
are compared with respect to the ones obtained through state-of-the-art
linear programming (LP) techniques \citep{hoang_steiner_2012} which
is able, for these particular instances, to find the optimal solutions.
The sign ``-'' denotes that no solution has been found. As shown
in table \ref{tab:Koch}, the gaps between the best energies achieved
by MS (in bold letters) and LP are always smaller than $4\%$ and
in two cases, for the multi-aligned \textit{augmenteddense-2} and
\textit{terminalintensive-2} instances, we reach the same performances
of LP, obtaining the optimal solutions. We stress that these graphs
are very loopy and far from being locally tree-like but nevertheless
we achieve good performances thanks to the reinforcement procedure
along with the introduction of the modified heuristics. Four examples of VLSI solutions are plotted in figure \ref{grids}.

It is worth noting that the greedy procedure (repeated for several
permutations of the order in which the trees were considered) fails
after few sequential searches. The average number of packed trees
$\left\langle M\right\rangle $ before the stop of the algorithm is
reported in the last column of table \ref{tab:Koch}. After these
greedy steps there exists one communication for which the connection
of all terminals is impossible: either the remaining graph, after
the pruning described in section \ref{subsec:Greedy-algorithm}, is
composed of disconnected components (one or more terminals are disconnected
from the rest of the network) or a possible connection may violate
the hard topological constraints.

\begin{table}
\begin{centering}
	\rotatebox{90}{
	\begin{tabular}{c|cccc|c|c|c|c|c|c}
 & Type & Size & $M$ & $T_{tot}$ & Heur. ``J'' & Heur. ``N'' & Rein. Max Sum & LP (opt) & Gap (MS, LP) \% & $\left\langle M\right\rangle $ Greedy\tabularnewline
\hline 
augmenteddense-2 & Multi-aligned & 16x18x2 & 19 & 59 & \textbf{504} (b) 506 (f) & 507 & 508 (f) \textbf{504} (b) & 504  & 0 \% & 14 (b) 2 (f)\tabularnewline
augmenteddense-2 & Multi-crossed & 16x18x2 & 19 & 59 & \textbf{503} & - & - & 498 & 1.0 \% & 7 (b) 5 (f)\tabularnewline
dense-3 & Multi-crossed & 15x17x3 & 19 & 59 & 487 & 488 & \textbf{485} & 464 & 4.0 \% & 11 (b) 7 (f)\tabularnewline
difficult-2 & Multi-aligned & 23x15x2 & 24 & 66 & \textbf{535} & 538 & 538 & 526 & 1.7 \% & 15 (b) 6 (f)\tabularnewline
difficult-2x & Multi-aligned & 23x15x2 & 24 & 66 & \textbf{560} & - & - & unknown &  & 14 (b) 4 (f)\tabularnewline
difficult-2y & Multi-aligned & 23x15x2 & 24 & 66 & \textbf{4776} & 4829 & 4816 & unknown &  & 14 (b) 8 (f)\tabularnewline
difficult-2z & Multi-aligned & 23x15x2 & 24 & 66 & \textbf{1060} & 1063 & 1061 & unknown &  & 16 (b) 7 (f)\tabularnewline
modifieddense-3 & Multi-crossed & 16x17x3 & 19 & 59 & \textbf{492} & 496 & 495 & 479 & 2.6 \% & 10 (b) 3 (f)\tabularnewline
moredifficult-2 & Multi-aligned & 22x15x2 & 24 & 65 & \textbf{542} & \textbf{542} & 546 & 522 & 3.8 \% & 13 (b) 5(f)\tabularnewline
pedabox-2 & Multi-aligned & 15x16x2 & 22 & 56 & \textbf{405} & \textbf{405} & \textbf{405} & 390 & 3.8 \% & 10 (b) 5 (f)\tabularnewline
terminalintensive-2 & Multi-aligned & 23x16x2 & 24 & 77 & \textbf{596} (f) 599 (b) & 617 & 620 & 596 & 0 \% & 13 (b) 5 (f)\tabularnewline
\end{tabular}
}
\par\end{centering}
\caption{Results for circuit layout instances \label{tab:Koch}}
\end{table}

\section{Summary of results}

Using Max-Sum algorithm, we have explored through simulations some
interesting theoretical questions in random graphs which we summarize
here. Simulations (up to $N=700$, or around $2\times10^{5}$ edges)
suggest that for the Steiner Tree packing problem on complete graphs
with uniform independent weights, the energy converges to a constant
value if the fraction of terminal vertices is kept constant, in agreement
with known results for single Steiner trees \citep{angel_sharp_2012}.

We have observed a non-negligible gap (up to 7\% in the solution energy
and increasing with tree depth) between a greedy solution (which is
commonly used by practitioners and consists in sequentially optimizing
each communication and removing its used components from the graph)
and the joint optimum computed by MS. Interestingly this gap is greatly
expanded (up to 80\% in experiments) with weights that are positively
correlated. For the edge-disjoint problem, we have compared all model
variants on random regular graphs with various parameters (degree,
number of terminals, number of trees), confirming the convenience
of each of them in a different parameter region. Simulations on regular
lattice graphs give qualitatively similar results.

Finally, we have attempted to optimize a set of publicly available
benchmark problems (including 3D tree packing problems), some of which
have known optimum. Results are encouraging, as the solutions provided
by the Max-Sum algorithm show a gap no larger than 4\% in all cases
(0\% in some cases) when the optimum is known. We expect this gap
to be generally independent of the problem size, which suggests that
this strategy could be extremely useful for large-scale industrial
problems.
\begin{acknowledgments}
We acknowledge Fondazione CRT for project SIBYL under the initiative
``La ricerca dei Talenti''. AB acknowledges funding by INFERNET, European Union's Horizon 2020 research
and innovation program under the Marie Sklodowska-Curie grant agreement No 734439,  and 
PRIN project 2015592CTH\_003 from the Italian ministry of university and research. We warmly thank F. Altarelli, R. Zecchina
for interesting discussions and T. Koch for providing us with the
VLSI instances.
\end{acknowledgments}

\bibliographystyle{plain}
\bibliography{multi_steiner}

\pagebreak{}

\appendix

\section{Message-Passing equations for V-DStP \label{sec:AppmesspasVertex}}

Consider the compatibility function for node $i$
\begin{equation}
\begin{aligned}\psi_{i}^{V}\left(\boldsymbol{d}_{i},\boldsymbol{\mu}_{i}\right)= & \prod_{k\in\partial i}\delta_{\mu_{ki},0}\delta_{d_{ki},0}+\sum_{\mu=1}^{M}\sum_{d>0}\sum_{k\in\partial i}\delta_{\mu_{ki},\mu}\delta_{d_{ki},-d}\prod_{l\in\partial i\setminus k}\left(\delta_{\mu_{li},\mu}\delta_{d_{li},d+1}+\delta_{\mu_{li},0}\delta_{d_{li},0}\right)+\\
 & +\sum_{\mu}\delta_{c_{i}^{\mu},0}\sum_{\substack{d>0}
}\sum_{k\in\partial i}\delta_{\mu,\mu_{ki}}\delta_{-d,d_{ki}}\sum_{l\in\partial i\backslash k}\delta_{\mu,\mu_{li}}\delta_{d_{li},d}\prod_{m\in\partial i\backslash\left\{ k,l\right\} }\delta_{\mu_{mi},0}\delta_{d_{mi},0}
\end{aligned}
\end{equation}

For sake of simplicity we split $\psi_{i}^{V}\left(\boldsymbol{d},\boldsymbol{\mu}\right)$
in
\begin{equation}
\psi_{i}^{V}\left(\boldsymbol{d}_{i},\boldsymbol{\mu}_{i}\right)=\psi_{i}^{(1)}\left(\boldsymbol{d}_{i},\boldsymbol{\mu}_{i}\right)+\psi_{i}^{(2)}\left(\boldsymbol{d}_{i},\boldsymbol{\mu}_{i}\right)+\psi_{i}^{(3)}\left(\boldsymbol{d}_{i},\boldsymbol{\mu}_{i}\right)
\end{equation}
where
\begin{eqnarray}
\psi_{i}^{(1)}\left(\boldsymbol{d}{}_{i},\boldsymbol{\mu}_{i}\right) & = & \prod_{k\in\partial i}\delta_{\mu_{ki},0}\delta_{d_{ki},0}\\
\psi_{i}^{(2)}\left(\boldsymbol{d}_{i},\boldsymbol{\mu}_{i}\right) & = & \sum_{\mu=1}^{M}\sum_{d>0}\sum_{k\in\partial i}\left[\delta_{\mu_{ki},\mu}\delta_{d_{ki},-d}\prod_{l\in\partial i\setminus k}\left(\delta_{\mu_{li},\mu}\delta_{d_{li},d+1}+\delta_{\mu_{li},0}\delta_{d_{li},0}\right)\right]\\
\psi_{i}^{(3)}\left(\boldsymbol{d}_{i},\boldsymbol{\mu}_{i}\right) & = & \sum_{\mu}\delta_{c_{i}^{\mu},0}\sum_{\substack{d>0}
}\sum_{k\in\partial i}\delta_{\mu,\mu_{ki}}\delta_{-d,d_{ki}}\sum_{l\in\partial i\backslash k}\delta_{\mu,\mu_{li}}\delta_{d_{li},d}\prod_{m\in\partial i\backslash\left\{ k,l\right\} }\delta_{\mu_{mi},0}\delta_{d_{mi},0}
\end{eqnarray}

Using $\psi_{i}^{(1)}\left(\boldsymbol{d}_{i},\boldsymbol{\mu}_{i}\right)$,
$\psi_{i}^{(2)}\left(\boldsymbol{d}_{i},\boldsymbol{\mu}_{i}\right)$
and $\psi_{i}^{(3)}\left(\boldsymbol{d}_{i},\boldsymbol{\mu}_{i}\right)$
inside \eqref{eq:bp1} we can compute the message $m_{ij}\left(d_{ij},\mu_{ij}\right)$
as a sum of three contributions, namely

$m_{ij}\left(d_{ij},\mu_{ij}\right)=m_{ij}^{(1)}\left(d_{ij},\mu_{ij}\right)+m_{ij}^{(2)}\left(d_{ij},\mu_{ij}\right)+m_{ij}^{(3)}\left(d_{ij},\mu_{ij}\right)$for

\begin{eqnarray}
m_{ij}^{(1)}\left(d_{ij},\mu_{ij}\right) & = & \sum_{\substack{\left\{ d_{ki},\mu_{ki}\right\} :\\
k\in\partial i\setminus j
}
}e^{-\beta\sum_{\mu}c_{i}^{\mu}\prod_{k\in\partial i}\left(1-\delta_{\mu_{ki},\mu}\right)}\prod_{l\in\partial i}\delta_{\mu_{li},0}\delta_{d_{li},0}\prod_{k\in\partial i\setminus j}n_{ki}\left(d_{ki},\mu_{ki}\right)\\
 & = & e^{-\beta\sum_{\mu}c_{i}^{\mu}}\delta_{\mu_{ij},0}\delta_{d_{ij},0}\prod_{k\in\partial i\setminus j}m_{ki}\left(0,0\right)\nonumber \\
m_{ij}^{(2)}\left(d_{ij,}\mu_{ij}\right) & = & \sum_{\mu}e^{-\beta\sum_{\mu}c_{i}^{\mu}\prod_{k\in\partial i}\left(1-\delta_{\mu_{ki},\mu}\right)}\sum_{d>0}\left\{ \delta_{d_{ji},-d}\delta_{\mu_{ji},\mu}\prod_{k\in\partial i\backslash j}\left[n_{ki}\left(d+1,\mu\right)+n_{ki}\left(0,0\right)\right]+\right.\\
 &  & \left.+\left(\delta_{d_{ji},d+1}\delta_{\mu_{ij},\mu}+\delta_{d_{ji},0}\delta_{\mu_{ji},0}\right)\sum_{k\in\partial i\backslash j}n_{ki}\left(-d,\mu\right)\prod_{l\in\partial i\backslash\{j,k\}}\left[n_{li}\left(d+1,\mu\right)+n_{li}\left(0,0\right)\right]\right\} \nonumber \\
m_{ij}^{(3)}\left(d_{ij},\mu_{ij}\right) & = & \sum_{\mu}\delta_{c_{i}^{\mu},0}\sum_{d>0}\left[\delta_{\mu,\mu_{ji}}\delta_{d_{ji},-d}\sum_{k\in\partial i\backslash j}n_{ki}\left(d,\mu\right)\prod_{l\in\partial i\backslash\left\{ j,k\right\} }n_{li}\left(0,0\right)+\right.\\
 &  & +\delta_{\mu,\mu_{ji}}\delta_{d_{ji},d}\sum_{k\in\partial i\backslash j}n_{ki}\left(-d,\mu\right)\prod_{l\in\partial i\backslash\left\{ j,k\right\} }n_{li}\left(0,0\right)+\nonumber \\
 &  & \left.+\delta_{\mu,\mu_{ji}}\delta_{d_{ji},0}\sum_{k\in\partial i\backslash j}n_{ki}\left(d,\mu\right)\sum_{l\in\partial i\backslash\left\{ j,k\right\} }n_{li}\left(-d,\mu\right)\prod_{m\in\partial i\backslash\left\{ k,l,j\right\} }n_{mi}\left(0,0\right)\right]\nonumber 
\end{eqnarray}

If now we use that $d_{ji}=-d_{ij}$ and $\mu_{ij}=\mu_{ji}$ we can
write the following set of equations:

\begin{eqnarray*}
m_{ij}\left(d,\mu\right) & = & \prod_{k\in\partial i\backslash j}\left[n_{ki}\left(d+1,\mu\right)+n_{ki}(0,0)\right]\\
 &  & \quad\quad+\delta_{c_{i}^{\mu},0}\sum_{k\in\partial i\backslash j}n_{ki}\left(d,\mu\right)\prod_{l\in\partial i\backslash\left\{ j,k\right\} }n_{li}\left(0,0\right)\quad\quad\forall d>0,\mu\neq0\\
m_{ij}\left(d,\mu\right) & = & \sum_{k\in\partial i\backslash j}n_{ki}\left(d+1,\mu\right)\prod_{l\in\partial i\backslash\{j,k\}}\left[n_{li}\left(d,\mu\right)+n_{li}(0,0)\right]\\
 &  & \quad\quad+\delta_{c_{i}^{\mu},0}\sum_{k\in\partial i\backslash j}n_{ki}\left(d,\mu\right)\prod_{l\in\partial j\backslash\left\{ j,k\right\} }n_{li}\left(0,0\right)\quad\quad\quad\forall d<0,\mu\neq0
\end{eqnarray*}

For $d=\mu=0$

\begin{eqnarray*}
m_{ij}\left(0,0\right) & = & e^{-\beta\sum_{\mu}c_{i}^{\mu}}\prod_{k\in\partial i\backslash j}n_{ki}\left(0,0\right)+\sum_{\mu\neq0}\sum_{d<0}\sum_{k\in\partial i\backslash j}n_{ki}\left(d+1,\mu\right)\prod_{l\in\partial i\backslash\{j,k\}}\left[n_{li}\left(d,\mu\right)+n_{li}(0,0)\right]+\\
 &  & +\sum_{\mu\neq0}\sum_{d<0}\sum_{k\in\partial i\backslash j}n_{ki}\left(d,\mu\right)\sum_{l\in\partial i\backslash\left\{ j,k\right\} }n_{li}\left(-d,\mu\right)\prod_{m\in\partial i\backslash\left\{ k,l,j\right\} }n_{mi}\left(0,0\right)
\end{eqnarray*}

\section{Recursive expression of $Z^{q}$ for the E-DStP \label{sec:RecursiveZ}}

From Eq. \eqref{eq:Zeta_M}

\begin{equation}
Z_{\mathbf{\boldsymbol{x}}}^{q}=\sum_{\substack{\boldsymbol{d}_{i},\boldsymbol{\mu}_{i}\\
\mu_{ki}\leq q,\,\forall k\in\partial i
}
}\psi_{i}^{E}\left(\boldsymbol{d}_{i},\boldsymbol{\mu}_{i}\right)e^{-\beta\sum_{\mu}c_{i}^{\mu}\prod_{k\in\partial i}\left(1-\delta_{\mu_{ki},\mu}\right)}\prod_{k\in\partial i}\mathbb{I}\left[x_{k}=1-\delta_{d_{ki},0}\right]n_{ki}\left(d_{ki},\mu_{ki}\right)
\end{equation}
we underline the possible contribution to a communication $q$ from
at least one on the neighbors of $i$ as

\begin{eqnarray*}
Z_{\mathbf{\boldsymbol{x}}}^{q} & = & \sum_{\substack{\boldsymbol{d},_{i}\boldsymbol{\mu}_{i}\\
\mu_{ki}\leq q
}
}\psi_{i}^{E}\left(\boldsymbol{d}_{i},\boldsymbol{\mu}_{i}\right)e^{-\beta\sum_{\mu}c_{i}^{\mu}\prod_{k\in\partial i}\left(1-\delta_{\mu_{ki},\mu}\right)}\times\\
 &  & \times\prod_{\substack{k\in\partial i:\\
\mu_{ki}=q
}
}\mathbb{I}\left[x_{k}=1\right]\mathbb{I}\left[d_{ki}\neq0\right]n_{ki}\left(d_{ki},\mu_{ki}\right)\prod_{\substack{k\in\partial i:\\
\mu_{ki}\leq q-1
}
}\mathbb{I}\left[x_{k}=1-\delta_{d_{ki},0}\right]n_{ki}\left(d_{ki},\mu_{ki}\right)
\end{eqnarray*}

Consider a vector $\boldsymbol{x}$ such that there exists at least
one component $x_{k}=1$ for $d_{ki}\neq0,\,\mu_{ki}=q$ and possibly
other components different from zero assigned to one of the possible
sub-graph $\mu\leq q-1$. This vector can be seen as the superposition
of all vectors $\boldsymbol{y}\leq\boldsymbol{x}$, that is, all vectors
having at most the same number of non-zeros of $\boldsymbol{x}$ and
the component $y_{k}=0$ each time $\mu_{ki}=q$; all remaining components
must satisfy $y_{k'}=1-\delta_{d_{k'i},0}$ for $\mu_{k'i}\leq q-1$.
Thus:
\begin{eqnarray}
Z_{\mathbf{\boldsymbol{x}}}^{q} & = & \sum_{\substack{\boldsymbol{d}_{i},\boldsymbol{\mu}_{i}\\
\mu_{ki}\leq q
}
}\sum_{\boldsymbol{y\leq\boldsymbol{x}}}e^{-\beta c_{i}^{q}\prod_{k\in\partial i}\left(1-\delta_{\mu_{ki},q}\right)}\psi_{i}^{q}\left(\boldsymbol{d}_{i},\boldsymbol{\mu}_{i}\right)\prod_{\substack{k\in\partial i:\\
y_{k}=0,\\
x_{k}=1
}
}n_{ki}\left(d_{ki},\mu_{ki}\right)\delta_{\mu_{ki},q}\times\\
 &  & \times\prod_{p\leq q-1}\prod_{\substack{k\in\partial i:\\
\mu_{ki}\leq q-1
}
}e^{-\beta c_{i}^{p}\prod_{k\in\partial i}\left(1-\delta_{\mu_{ki},p}\right)}\psi_{i}^{p}\left(\boldsymbol{d}_{i},\boldsymbol{\mu}_{i}\right)\mathbb{I}\left[y_{k}=1-\delta_{d_{ki},0}\right]n_{ki}\left(d_{ki},\mu_{ki}\right)
\end{eqnarray}
where we have made use of the expression of $\psi_{i}^{E}$ in \eqref{eq:psi-edge-disj}.
If we now collect the sum over $\boldsymbol{y}\leq\boldsymbol{x}$
and we explicitly use the constraints on depth and communication variables
we find

\begin{eqnarray}
Z_{\boldsymbol{x}}^{q} & = & \sum_{\boldsymbol{y}\leq\boldsymbol{x}}\left\{ \sum_{\substack{\boldsymbol{d}_{i},\boldsymbol{\mu}_{i}\\
\mu_{ki}\leq q
}
}e^{-\beta c_{i}^{q}\prod_{k\in\partial i}\left(1-\delta_{\mu_{ki},q}\right)}\psi_{i}^{q}\left(\boldsymbol{d},\boldsymbol{\mu}\right)\prod_{\substack{k\in\partial i\\
y_{k}=0\\
x_{k}=1
}
}n_{ki}\left(d_{ki},\mu_{ki}\right)\delta_{\mu_{ki},q}\times\right.\\
 &  & \left.\times\sum_{\substack{\boldsymbol{d}_{i},\boldsymbol{\mu}_{i}\\
\mu_{ki}\leq q-1
}
}\prod_{p\leq q-1}e^{-\beta c_{i}^{p}\prod_{k\in\partial i}\left(1-\delta_{\mu_{ki},p}\right)}\psi_{i}^{p}\left(\boldsymbol{d},\boldsymbol{\mu}\right)\prod_{\substack{k\in\partial i}
}\mathbb{I}\left[y_{k}=1-\delta_{d_{ki},0}\right]n_{ki}\left(d_{ki},\mu_{ki}\right)\right\} \\
 & = & \sum_{\boldsymbol{y}\leq\boldsymbol{x}}\left(g_{\boldsymbol{y}}^{0}+g_{\boldsymbol{y}}^{b}+g_{\boldsymbol{y}}^{f}\right)Z_{\boldsymbol{y}}^{q-1}
\end{eqnarray}
where 
\begin{eqnarray*}
g_{\boldsymbol{y}}^{0} & = & e^{-\beta c_{i}^{q}}\prod_{\substack{k\in\partial i\\
y_{k}=0\\
x_{k}=1
}
}n_{ki}\left(0,0\right)\\
g_{\boldsymbol{y}}^{b} & = & \sum_{d>0}\sum_{\substack{j\in\partial i\\
y_{j}=0\\
x_{j}=1
}
}n_{ji}\left(-d,q\right)\prod_{\substack{k\in\partial i\setminus j\\
y_{k}=0\\
x_{k}=1
}
}\left[n_{ki}\left(d+1,q\right)+n_{ki}\left(0,0\right)\right]\\
g_{\boldsymbol{y}}^{f} & = & \delta_{c_{i}^{q},0}\sum_{d>0}\sum_{\substack{j\in\partial i\\
y_{j}=0\\
x_{j}=1
}
}n_{ji}\left(-d,q\right)\sum_{\substack{k\in\partial i\backslash j\\
y_{k}=0\\
x_{k}=1
}
}n_{ki}\left(d,q\right)\prod_{\substack{l\in\partial i\setminus\left\{ j,k\right\} \\
y_{l}=0\\
x_{l}=1
}
}n_{li}\left(0,0\right)
\end{eqnarray*}

In the special case in which no communications is flowing within the
graph, that is for $q=0$, we must impose the value of $Z_{\mathbf{x}}^{0}$
through

\begin{equation}
Z_{\mathbf{x}}^{0}=e^{-\beta\sum_{\mu}c_{i}^{\mu}}\mathbb{I}\left[\boldsymbol{x}=\boldsymbol{0}\right]\prod_{j\in\partial i}n_{ji}\left(0,0\right)
\end{equation}

\section{From E-DStP to a weighted maximum matching problem\label{sec:From-E-DStP-to}}

Let us explicit the dependency on $\boldsymbol{\mu}_{i}$ of \eqref{eq:Psi-Ed-match} 

\begin{eqnarray}
\psi_{i}^{E}\left(\boldsymbol{d}_{i},\boldsymbol{\mu}_{i}\right) & = & \sum_{\mathbf{s}}\left\{ \prod_{\mu:s_{\mu}>0}\sum_{k\in\partial i}\delta_{\mu_{ki},\mu}\delta_{d_{ki},-s_{\mu}}\prod_{l\in\partial i\setminus k}\left[\delta_{\mu_{li},\mu}\delta_{d_{li},s_{\mu}+1}+\left(1-\delta_{\mu_{li},\mu}\right)\right]\right.+\\
 &  & \left.+\prod_{\mu:s_{\mu}=0}\prod_{k\in\partial i}\left(1-\delta_{\mu_{ki},\mu}\right)\delta_{d_{ki},s_{\mu_{ki}}}\right\} 
\end{eqnarray}

and let us introduce a partial partition function as in \eqref{eq:Zeta_i_x}and
let us underline the $\boldsymbol{s}-$dependence as

\begin{eqnarray}
Z_{i} & = & \sum_{\boldsymbol{d},\boldsymbol{\mu}_{i}}\psi_{i}^{E}\left(\boldsymbol{d}_{i},\boldsymbol{\mu}_{i}\right)e^{-\beta\sum_{\mu}c_{i}^{\mu}\prod_{k\in\partial i}\left(1-\delta_{\mu_{ki},\mu}\right)}\prod_{k\in\partial i}n_{ki}\left(d_{ki},\mu_{ki}\right)\\
 & = & \sum_{\mathbf{s}}Q_{\mathbf{s}}
\end{eqnarray}
where the function $Q_{\boldsymbol{s}}$reads (here we collect the
topological constraints in $f_{k\mu}$ for a neighbor $k\in\partial i$
participating in sub-graph $\mu$)
\begin{equation}
Q_{\mathbf{s}}=\sum_{\boldsymbol{d}_{i}}\sum_{\left\{ \mu_{ki}:s_{\mu_{ki}}>0\vee\mu_{ki}=0\right\} }\prod_{k\in\partial i}n_{ki}\left(d_{ki},\mu_{ki}\right)\left\{ \prod_{\mu:s_{\mu}>0}e^{-\beta\sum_{\nu\neq\mu}c_{i}^{\nu}\prod_{k\in\partial i}\left(1-\delta_{\mu_{ki},\nu}\right)}\sum_{k\in\partial i}f_{k\mu}+e^{-\beta\sum_{\mu}c_{i}^{\mu}}\prod_{k\in\partial i}\delta_{\mu_{ki},0}\right\} 
\end{equation}

\begin{equation}
f_{k\mu}=\delta_{\mu_{ki},\mu}\delta_{d_{ki},-s_{\mu}}\prod_{l\in\partial i\setminus k}\left[\delta_{\mu_{li},\mu}\delta_{d_{li},s_{\mu}+1}+\left(1-\delta_{\mu_{li},\mu}\right)\right]
\end{equation}

Let us concentrate in the computation of $Q_{\mathbf{s}}$ for a fixed
$\mathbf{s}$. For simplicity of notation, we will assume, unless
explicitly noted, that $\mu$ indices run over the set $\left\{ \mu:s_{\mu}>0\right\} $.
Now as $f_{k\mu}f_{k\nu}=0$ if $\mu\neq\nu$ (because $\delta_{\mu_{ki},\mu}\delta_{\mu_{ki},\nu}=0$),
we have that

\begin{equation}
\delta_{\mu_{ki},\mu}\delta_{d_{ki},-s_{\mu}}\left[\delta_{\mu_{ki},\nu}\delta_{d_{ki},s_{\nu}+1}+\left(1-\delta_{\mu_{ki},\nu}\right)\right]=\delta_{\mu_{ki},\mu}\delta_{d_{ki},-s_{\mu}}
\end{equation}
and equivalently

\begin{equation}
\prod_{\nu}\left[\delta_{\mu_{ki},\nu}\delta_{d_{ki},s_{\nu}+1}+\left(1-\delta_{\mu_{ki},\nu}\right)\right]=\sum_{\nu}\delta_{\mu_{ki},\nu}\delta_{d_{ki},s_{\nu}+1}+\delta_{\mu_{ki},0}\delta_{d_{ki},0}
\end{equation}

Thus 
\[
\prod_{\mu}\sum_{k\in\partial i}f_{k\mu}=\sum_{\pi}\prod_{\mu}f_{\pi_{\mu}\mu}
\]
where the sum $\sum_{\pi}$ runs over all the possible coupling between
communications and neighbors of node $i$. Mathematically we have
defined the one-to-one functions $\pi$ 
\[
\pi:\left\{ \mu:s_{\mu}>0\right\} \to\partial i
\]
with $\pi:\mu\mapsto\pi_{\mu}$. In the following, we will switch
to an alternative representation of functions $\pi$. If we denote
by $t_{k\mu}=\delta_{k,\pi_{\mu}}$, for a fixed $\pi$ we obtain
\begin{eqnarray*}
\prod_{\mu}f_{\pi_{\mu}\mu} & = & \prod_{\mu}\delta_{\mu_{\pi_{\mu}i},\mu}\delta_{d_{\pi_{\mu}i},-s_{\mu}}\prod_{l\in\partial i\setminus\pi_{\mu}}\left[\delta_{\mu_{li},\mu}\delta_{d_{li},s_{\mu}+1}+\left(1-\delta_{\mu_{li},\mu}\right)\right]\\
 & = & \prod_{k\in\partial i}\left(\sum_{\nu}\delta_{\mu_{ki},\nu}\delta_{d_{ki},s_{\nu}+1}+\delta_{\mu_{ki},0}\delta_{d_{ki},0}\right)^{1-\sum_{\nu}t_{k\nu}}\prod_{\mu}\left(\delta_{\mu_{ki},\mu}\delta_{d_{ki},-s_{\mu}}\right)^{t_{k\mu}}
\end{eqnarray*}
with the convention that $0^{0}=1$. Note that the vector $\boldsymbol{t}$
and the function $\pi$ contain the same information: we have that
$\sum_{k\in\partial i}t_{k\mu}=1-\delta_{s_{\mu},0}$ for each $\mu$
and $\sum_{\mu}t_{k\mu}\leq1$ for each $k\in\partial i$. These two
conditions are complete; for a vector $\boldsymbol{t}$ that satisfies
these two constraints, the corresponding function $\pi$ can be defined
naturally. We will have then

\begin{eqnarray}
Z_{i} & = & \sum_{\boldsymbol{s}}\sum_{\boldsymbol{t}}\prod_{\mu}e^{-\beta c_{i}^{\mu}\prod_{k\in\partial i}\left(1-\delta_{\mu_{ki},\mu}\right)}\mathbb{I}\left[\sum_{k\in\partial i}t_{k\mu}=1-\delta_{s_{\mu},0}\right]\prod_{k\in\partial i}\mathbb{I}\left[\sum_{\mu}t_{k\mu}\leq1\right]\times\\
 &  & \times\prod_{k\in\partial i}\left[\sum_{\nu}n_{ki}\left(s_{\nu}+1,\nu\right)+n_{ki}\left(0,0\right)\right]^{1-\sum_{\nu}t_{k\nu}}\prod_{\mu}n_{ki}\left(-s_{\mu},\mu\right)^{t_{k\mu}}\\
 & = & \sum_{\boldsymbol{s}}R_{\boldsymbol{s}}Z_{\boldsymbol{s}}
\end{eqnarray}

\end{document}